\documentclass[
reprint,
superscriptaddress,
floatfix,
citeautoscript,
amsmath,
amssymb,
prmaterials
]{revtex4-2}

\usepackage{graphicx}
\usepackage{dcolumn}
\usepackage{bm}
\usepackage{siunitx}
\usepackage{booktabs}
\usepackage{xcolor}
\usepackage{comment}
\usepackage{hyperref}

\newcommand{\PS}[2]{\{\SI{#1}{W}, \SI{#2}{mm/s}\}}

\begin{document}
\newcolumntype{C}[1]{>{\centering\let\newline\\\arraybackslash\hspace{0pt}}m{#1}}

\title{Local laser-induced solid-phase recrystallization of phosphorus-implanted Si/SiGe heterostructures for contacts below 4.2\,K}

\author{Malte Neul}
\author{Isabelle V. Sprave}
\affiliation{JARA-FIT Institute for Quantum Information, Forschungszentrum J\"ulich GmbH and RWTH Aachen University, Aachen, Germany}
\author{Laura K. Diebel}
\author{Lukas G. Zinkl}
\affiliation{Fakult\"at f\"ur Physik, Universit\"at Regensburg, 93040~Regensburg, Germany}
\author{Florian Fuchs}
\affiliation{Fraunhofer-Institut f\"ur Lasertechnik ILT, 52074~Aachen, Germany}
\author{Yuji Yamamoto}
\affiliation{IHP-Leibniz-Institut f\"ur Innovative Mikroelektronik, Im Technologiepark 25, 15236 Frankfurt (Oder), Germany}
\author{Christian Vedder}
\affiliation{Fraunhofer-Institut f\"ur Lasertechnik ILT, 52074~Aachen, Germany}
\author{Dominique Bougeard}
\affiliation{Fakult\"at f\"ur Physik, Universit\"at Regensburg, 93040~Regensburg, Germany}
\author{Lars R. Schreiber}
\email{lars.schreiber@physik.rwth-aachen.de}
\affiliation{JARA-FIT Institute for Quantum Information, Forschungszentrum J\"ulich GmbH and RWTH Aachen University, Aachen, Germany}
\affiliation{ARQUE Systems GmbH, 52074~Aachen, Germany}

\date{\today}
\begin{abstract}
Si/SiGe heterostructures are of high interest for high mobility transistor and qubit applications, specifically for operations below \SI{4.2}{K}. In order to optimize parameters such as charge mobility, built-in strain, electrostatic disorder, charge noise and valley splitting, these heterostructures require Ge concentration profiles close to mono-layer precision. Ohmic contacts to undoped heterostructures are usually facilitated by a global annealing step activating implanted dopants, but compromising the carefully engineered layer stack due to atom diffusion and strain relaxation in the active device region. We demonstrate a local laser-based annealing process for recrystallization of ion-implanted contacts in SiGe, greatly reducing the thermal load on the active device area. To quickly adapt this process to the constantly evolving heterostructures, we deploy a calibration procedure based exclusively on optical inspection at room-temperature. We measure the electron mobility and contact resistance of laser annealed Hall bars at temperatures below \SI{4.2}{K} and obtain values similar or superior than that of a globally annealed reference samples. This highlights the usefulness of laser-based annealing to take full advantage of high-performance Si/SiGe heterostructures.
\end{abstract}
\maketitle

\section{Introduction}
\label{sec:Intro}

The versatility of band and strain engineering within heterostructures makes them compelling candidates for use in cryogenic transistors such as the high electron mobility transistor (HEMT) \cite{Nazir2022, Zeng2022} and the heterojunction bipolar transistor (HBT) \cite{Najafizadeh2006, Ward2006, Hollmann2018, Hosseini2021}, as well as for qubit applications \cite{Yoneda2018, Noiri2022, Xue2022, Struck2023, Xue2023} that benefit from high carrier mobility, low electrostatic disorder and low electrical noise. These devices often aim to minimize scattering sites and charge noise by utilizing undoped heterostructures \cite{Maune2012}. Consequently, they require local doping to generate Ohmic contacts to the semiconductor e.g. to the buried conduction layer of a Si/SiGe quantum well. However, this necessitates high temperature annealing steps, approaching or even surpassing the growth temperature. This can lead to atom diffusion and intermixing at the interfaces \cite{Bougeard}. In addition, the elevated temperature leads to the propagation of misfit dislocations through the quantum well, reducing the tensile stress and increasing the amount of scatter sites \cite{Liu2022b}.

As local interface fluctuations become increasingly relevant and engineering precision at the atomic scale becomes a requirement \cite{Losert2023}, global annealing becomes a limitation, as its thermal impact potentially degrades the active region. This is especially relevant for strain engineered structures such as silicon/silicon-germanium (Si/SiGe) grown by molecular-beam-epitaxy (MBE) \cite{Liu2022} as well as qubit devices \cite{McJunkin2021, McJunkin2022, PaqueletWuetz2022, Losert2023} based on Si/SiGe heterostructures. There, sharp interfaces are required for lifting the out-of-plane conduction band minima degeneracy (valley splitting) in the tensile strained Si quantum well. Recently, Ge concentration profiles with monolayer resolution are introduced into the quantum well to compensate for the inaccessibility of atomistical abrupt changes in alloy concentration \cite{McJunkin2021}. The advantage of having these layers becomes significantly reduced if they are smeared out due to thermally activated diffusion. Insufficient valley splitting is thereby considered as the dominant limitation for scaling up Si/SiGe quantum computers, as local minima can lead to fast decoherence channels hampering coherent shuttling and operation at elevated temperature \cite{Vandersypen2017, Chen2021, Langrock2022}. Ohmic contacts are required for these quantum computing applications, the operation temperature of which  is below the temperature of liquid helium (\SI{4.2}{K}). This places a lower limit on the annealing temperature, as incomplete ionization, caused by freezing-out \cite{Beckers2020} has to be accounted for. This is critical as it can lead to a breakdown of conductivity in the contacts, reducing the signal-to-noise ratio and limiting the measurement bandwidth. Insufficient activation can even render the quantum well layer becoming electrically inaccessible altogether.

Recrystallization through solid-phase-epitaxy (SPE) in Si starts at temperatures exceeding \SI{550}{\degree C} \cite{Rimini1995}. However, practical constraints on annealing times, coupled with the need to prevent freezing-out, often necessitate to include a global heating step of \SI{700}{\degree C} and above in typical processes \cite{Lawrie2020,Seidler2022}.
Laser annealing (LA) was previously studied in ion-implanted Si to improve sheet resistance \cite{Taiwo2022}. It was also proposed as a solution to form surface-localized junctions for monolithically stacked devices to prevent degradation of lower laying structures \cite{Tabata2022, Tabata2022a}. Outside of dopant activation, laser annealing was used for relaxing SiGe layers grown on silicon on insulator (SOI) as virtual substrate for strained Si channels \cite{Mishima2005} and for frequency tuning of superconducting qubits by local oxide thickness control \cite{Kim2022}. However, no study was done on LA activated contacts in Si/SiGe, specifically for cryogenic use below \SI{4.2}{K}.

In this work, we demonstrate activation of implanted phosphor donors in an undoped Si/SiGe heterostructure by local annealing with a continuous-wave focused laser beam. The area, which is scanned by the laser, makes an excellent Ohmic contact to a buried Si/SiGe quantum well operating down to 1.6\,K and our LA process keeps the silicon surface topography intact for further device fabrication. We benchmark the LA performance by gated Si/SiGe Hall-bar. The contact resistance and electron mobility is at least equal compared to a globally annealed reference Hall-bar. We find that the laser power is the most critical process parameter, which depends on the details of the Si/SiGe heterostructure. We suggest a fast calibration method based on optical inspection after LA. Thus, LA is a viable alternative that can reduce the thermal budget and improve the quality of a Si/SiGe heterostructure in the active region, which is particularly interesting for SiGe quantum chips.

\section{Materials and methods}
\label{sec:Material_Methods}
\begin{figure*}
\includegraphics{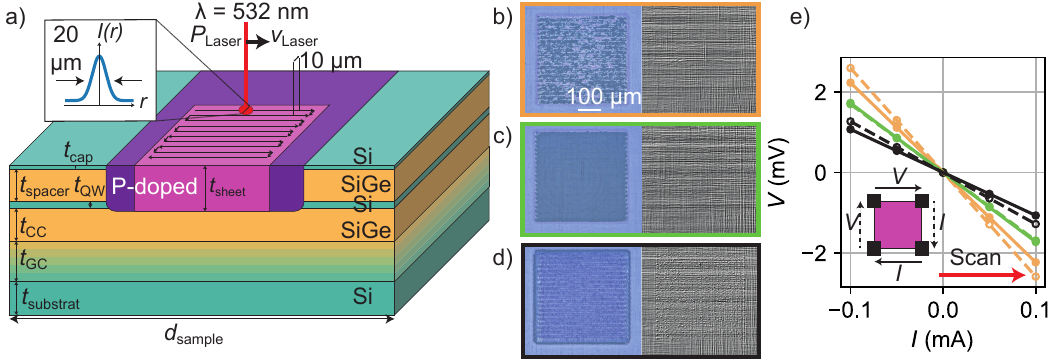}
\caption{\label{fig:sample_method} 
Laser annealing and process evaluation: a) Schematic of studied heterostructure (stack detailed in Tab. \ref{tab:Wafer_List}) and laser annealing setup. A CW laser with $\lambda$ = \SI{532}{nm} is scanned line-by-line (black line) over the sample and controlled via a shutter to only hit the designated area (pink area) inside the implanted regions (purple area). The tool allows control of the laser power $P$, scan speed $v$ and line spacing $\Delta y$. b) Optical inspection (left) and c-DIC image (right) of a $\SI{400}{\micro m} \times \SI{400}{\micro m}$ squares scanned with \PS{2.6}{25}, c) \PS{2.6}{2.5} and d) \PS{6.2}{2.5} on wafer B. e) Room-temperature current $I$ versus voltage $V$ characteristic for the squares shown in panels b)-d) (dot colors match frame colors). Data is obtained in van-der-Pauw geometry (black squares are electrical contacts in insert). Lines are linear least-square fits to the data. $I$ is measured parallel to the laser scanline (filled dots, solid line) and perpendicular to it (open dots, dashed line) as sketched in the insert.}
\end{figure*}
\subsection{Samples}
\label{subsec:Samples}

All four investigated heterostructures follow the same scheme with differences in layer thicknesses as shown in Fig. \ref{fig:sample_method} (a). A Si$_{1-x}$Ge$_x$ virtual substrate (VS), consisting of a graded-composition (GC) buffer with either step-wise or linearly increasing Ge content up to $x$ and an additional constant-composition (CC) buffer was deposited onto a Si substrate. This is followed by the growth of the active thin Si layer on the VS and then a second Si$_{1-x}$Ge$_x$ spacer layer. The Si layer is tensile strained, since it is thinner than the critical thickness for plastic relaxation. Finally, the spacer is capped by a thin Si layer to prevent direct oxidation of the Si$_{1-x}$Ge$_x$. The wafers were grown using molecular-beam-epitaxy (MBE) or chemical-vapor-deposition (CVD) and all details of the four wafers (A-D) are summarized in Tab. \ref{tab:Wafer_List}. All wafers were diced into $\SI{10}{mm}\,\times \SI{10}{mm}$ samples.
\begin{table}[]
\caption{Overview of the layer stack and growth conditions of the studied wafers.}
\label{tab:Wafer_List}
\begin{tabular}{@{}l|cccc@{}}
\toprule
Wafer                                                            & A      & B        & C        & D      \\ \midrule
$x$                                                              & 0.30   & 0.33     & 0.33     & 0.30   \\
$t_\mathrm{cap}$ (nm)                                            & 5.0    & 5.0      & 5.0      & 1.5    \\
$t_\mathrm{spacer}$ (nm)                                            & 30     & 35       & 55       & 45     \\
$t_\mathrm{QW}$ (nm)                                             & 10     & 10       & 10       & 10     \\
$t_\mathrm{CC}$ (\si{\micro m})                                  & 2.0    & 1.75     & 1.75     & 0.70   \\
$t_\mathrm{GC}$ (\si{\micro m})                              & 3.0    & 0.25     & 0.25     & 3.75   \\
Grading type                                                      & Linear & Step     & Step     & Linear \\
$t_\mathrm{substrate}$(\si{\micro m})                            & 725    & 725      & 725      & 425    \\
Growth method                                                    & CVD    & CVD      & CVD      & MBE    \\ \bottomrule
\end{tabular}
\end{table}
After an RCA cleaning step (details see App. \ref{app:sec:fab}, applied to wafer A-C, wafer D was excluded from this clean, as this would have resulted in complete removal of the Si capping layer), all samples were ion-implanted by a \SI{20}{kV} accelerated phosphorus beam with a flux of \SI{5E15}{cm}$^{-2}$ hitting the sample at an angle of \SI{7}{\degree} to its normal. From \textit{SRIM 2013} simulations \cite{Bromley1985}, a concentration above \SI{1E20}{\per \cubic \centi \meter} is expected up to a depth of \SI{64}{\nano \meter}, with the maximum concentration at \SI{26}{\nano \meter} below the surface.

\subsection{Setup}
\label{subsec:Setup}
LA is conducted by a custom-made \textit{MICROTECH LaserWriter, LW532 model}. It is equipped with a solid-state continuous-wave $\lambda = $ \SI{532}{nm} laser with a variable power $P$ between \SI{0.2}{W} and \SI{8}{W}. Its wavelength was chosen based on simulations of the heterostructure for high absorption and low transmission/reflectivity. The laser is focused down to a spot diameter of $\SI{20}{\um}$ with a Gaussian profile. The sample is placed on a ceramic chuck mounted on top of a movable xy-stage, the velocity of which can be varied between \SI{0.1}{mm \per s} and \SI{25.0}{mm \per s} along its fast axis. Pattern alignment and in-process monitoring are done via a CCD camera positioned within the optical path.
For operation, the laser spot is scanned back and forth over the sample along the fast axis with the scan velocity $v$ and is stepped after each line along the slow axis with a fixed distance $\Delta y = \SI{10}{\um}$ as indicated in Fig. \ref{fig:sample_method} (a) (detailed discussion see App. \ref{app:sec:steps}). To account for the inertia of the stage, the scan starts before the sample area for each line to ensure enough travel distance to reach the set scan velocity. The laser is controlled by a shutter with a reaction time of \SI{10}{ms}, which blocks the laser during scanning to prevent heating outside the designated areas and transients if laser output power was switched.

\subsection{Optical and electrical characterization (RT)}
\label{subsec:RT_Chara}

In the calibration of the laser annealing process, it is essential to monitor the level of recrystallization, alterations in surface topography and resulting electrical resistance for changing sets of process parameters. These criteria serve as crucial indicators of the success of the annealing step. Ideally, the calibration method should be fast and easily applicable, so that a large parameter space can be covered for each new heterostructure. To this end, it should involve minimal additional fabrication steps or added complexity. To achieve this, we utilize the fact that implantation damage induces a change in color, stemming from the difference in optical properties between crystalline and amorphous semiconductors \cite{Wesch1980}. Consequently, we can assess the degree of recrystallization by evaluating the extent of color change following the annealing process. This assessment can be done quickly by optical inspection. We examine the impact of LA on the surface topography by circular polarized light–differential interference contrast (c-DIC) \cite{Danz2004} imaging.

With this characterisation in place, we study the dependence of total regrown thickness on laser power $P$, the dwell time of the laser (which is inversely related to the scan velocity $v$) and the thermal conductivity/capacity of the surrounding material. For this, we anneal multiple squares with an edge length of $\SI{400}{\um}$ with different parameter sets $\{P, v\}$ on samples of each wafer. After LA, optical and c-DIC images are taken under a microscope. Depending on the set of parameters employed, we observe varying degrees of color change in the annealed areas within the optical images. We cluster these into three categories: for \{low $P$, high $v$\} we observe an inhomogeneous color change as shown in Fig. \ref{fig:sample_method} (b) (orange). The color change occurs preferentially in the middle of the scanned line. At this point, the only surface topology we observe in the samples corresponds to the typical cross-hatching pattern commonly found on the surface of Si/SiGe heterostructures \cite{Chen2002}. The proportion of changed regions increases with \{increasing $P$, decreasing $v$\} until a homogeneous color change is reached (Fig. \ref{fig:sample_method} (c) (green)), without an observable change in surface topology. When power (velocity) is further increased (decreased), bright spots appear in the optical image towards the middle of the scan line, shown in Fig. \ref{fig:sample_method} (d) (black). The spots are also visible in the c-DIC images. 

Presumably, the initial temperature in the sample does not exceed the melting point. Therefore, regrowth takes place via solid-phase-epitaxy (SPE) from the bottom up and increases in speed according to a Arrhenius equation for higher temperatures \cite{Marine1983, Boyd1983}. Thus, we attribute the inhomogeneous color change to a threshold range, where certain parts of the intensity distribution of the laser are already sufficient to heal the implanted damage over the full depth, while the remaining spots are only partly recrystallized (see Fig. \ref{fig:sample_method} (b)). The homogeneous color change (Fig. \ref{fig:sample_method} (c)) indicates then that the entire implanted area has fully healed. 
Upon further laser power increase, we observe a change in surface topography as indicated by the appearance of the black dots in the c-DIC image (as seen in Fig. \ref{fig:sample_method} (d)).
We speculate that this indicates local exceeding of the heterostructure's melting point of approximately \SI{1220}{^\circ C} \cite{Olesinski1984}. At this point, a liquid phase locally develops, the volume of which extends further down as the LA time is increased \cite{Marine1983}. After cooling below the melting point, the liquid phase crystallizes, with the underlying semiconductor serving as a seed. We aim not to change the surface during the process in order to keep further fabrication steps consistent. Consequently, we dismiss parameter sets, where optical inspection indicates surface degradation.

In order to qualify the interpretation of optical inspection, we evaluate the electrical conductivity of the annealed areas first at \SI{300}{K}. We measure the electrical resistivity of the three annealed squares in a four-probe van-der-Pauw (vdP) geometry at room temperature (RT) as shown color coded in Fig. \ref{fig:sample_method} (e). To this end we evaporate \SI{150}{nm} thick aluminium contact pads onto the corners of each square. We conduct the experiment twice, once along the fast scan direction of the laser beam (indicated by filled dots and solid line) and once perpendicular to it (indicated by open dots and dashed line), and apply a linear fit to each data set. We observe a linear and thus Ohmic behavior for all measured squares and orientations. The extracted resistance decreases from the inhomogeneous (orange) to the homogeneous (green) to the surface degradation case (black). Notably, for the inhomogeneous and degradation squares, the resistance displays a dependence on the measurement orientation, with lower resistances observed when measured along the laser scanline. This distinction is not observed in the case of the homogeneous sample. The decline in resistance matches expectations, considering that the ratio of electrically active dopants has been observed to increase with temperature \cite{Luong2013} and our optical study suggests that the temperature increases between each of the squares. Moreover, this temperature variation becomes more pronounced if the sample surpass the melting point, as assumed in the surface degradation square. The reason for this is that the diffusion rate in the liquid phase significantly exceeds that in the solid phase \cite{Boyd1983}.
This can explain the observed preferential direction of the resistance, since the temperature is highest towards the center of the laser. Conversely, perpendicular to the scan direction, the temperature minimum is found between two lines. In the inhomogeneous case, this implies the presence of remaining narrow amorphous regions which enhance the resistance along the current direction. On the other hand, in the surface degradation case, the melting predominantly occurs toward the laser center, generating areas with a high activation ratio interleaved with unmolten regions perpendicular to the scanline. In contrast, the homogeneous case achieves a consistent activation ratio without the presence of any amorphous or molten resistance elements. The electrical measurements validate our interpretation of the optical inspection in terms of the stage of recrystallization.

As such, optical inspection provides a rapid and effective means of assessing the feasibility of a parameter set $\{P, v\}$ for a given heterostructure. We define a successful parameter set by the presence of a uniform color change without any discernible surface morphology alterations. This evaluation technique enables efficient calibration of the LA process for each new heterostructure.

\section{Calibration results}
\label{sec:Callibration}
\begin{figure*}
\includegraphics{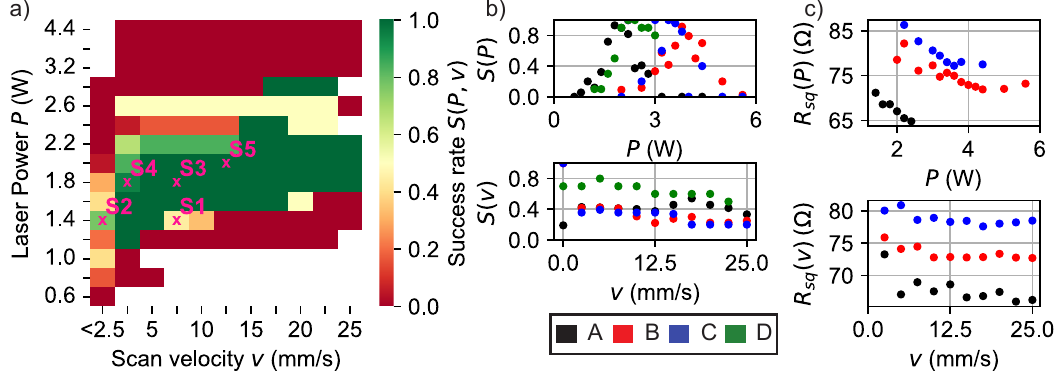}
\caption{\label{fig:RT_calibration} Process calibration: a) Map of success rate $S$ as a function of parameter set $\{P,v\}$ for wafer A. A successful parameter set is defined as exhibiting a homogeneous color change and not indication of surface morphology change. b) Projected success rates $S(P)$ (top) and $S(v)$ (bottom) obtained by averaging along the $v$ or $P$ axis for wafers A-D (color coded). c) Projected sheet resistances $R_\mathrm{sq}(P)$ (top) and $R_\mathrm{sq}(v)$ (bottom) of successfully annealed regions obtained by vdP measurements at \SI{300}{K}.}
\end{figure*}


Building upon the efficient determination of successful parameter sets through optical inspection, we focus on investigating how these parameter sets are influenced by changes in the heterostructure. This investigation is critical, as continuous optimization of various aspects of the Si/SiGe layer stacks necessitates adaptations in the laser annealing process. We systematically vary laser power $P$ and scan velocity $v$ to anneal squares onto samples of all four wafers and optically inspect the results. Several parameter sets are studied multiple times to assess process stability. We define the success rate $S$ for a given parameter set $\{P, v\}$ as the number of successful squares in relation to the total number of annealed squares. A map of the $S$ is shown for wafer A in Fig. \ref{fig:RT_calibration} (a). The maps for wafers B-D are shown in App. \ref{app:sec:optic}. A parameter window with high success rates is observed (green area). This high success parameter window is narrow in $P$ and broad in $v$. Thus, the laser power is a more critical parameter for LA of wafer A. To compare the successful parameter windows among the wafers, we calculate projections of average success rates $S(P)$ ($S(v)$) along the $v$ ($P$) axis, given by
\begin{equation}
    S(i) = \frac{1}{N_{j}}\sum_{j} S(P,v) \text{ with } i\neq j \in [v,P],
\end{equation}
where $N_{j}$ are the number of tested parameter sets along each projection. The result are shown in Fig. \ref{fig:RT_calibration} (b). The $P$-projection (top) shows a dependence of the success rate on laser power visible as peak. The position of this peak varies for the four wafers  studied, with a maximum distance between wafer A and B. The success rate maximum of wafer C is closer to the B-peak and the D-peak close to the one of A. The $v$-projection (bottom) shows no systematic change of $S(v)$ for all four wafers, which confirms that the scanning velocity is uncritical for all considered samples. The laser power, however, needs recalibration for all wafers.  Note that we observed negligible change in the reflection coefficient of the laser wavelength among samples. 

For wafers A, B, and C, we conduct vdP measurements within each successfully annealed region. In Fig. \ref{fig:RT_calibration} (c), sheet resistance projections $R_\mathrm{sq}(P)$ ($R_\mathrm{sq}(v)$), obtained by employing the previously described method along $v$ ($P$), are shown. We observe a trend of decreasing resistance projection for increasing $P$, while the projections are constant in $v$. We obtain minimum sheet resistances of $R_\mathrm{sq,A} = \SI{64.4}{\ohm}$, $R_\mathrm{sq,B} = \SI{70.1}{\ohm}$ and $R_\mathrm{sq,C} = \SI{74.4}{\ohm}$.
\begin{table}[]
\caption{Minimum sheet resistances that are achieved with a successful parameter set for each wafer examined. Corresponding values for power $P$ and scanning velocity $v$ are given.}
\label{tab:CalibrationSummary}
\begin{tabular}{@{}C{1.75cm}|C{1.75cm}C{1.75cm}C{1.75cm}@{}}
\toprule
Wafer & $R_\mathrm{sq,min}$ (\si{\Omega}) & $P$  (\si{W}) & $v$ \, (\si{mm/s}) \\ \midrule
A     & 64.4                          & 2.4          & 15.0          \\
B     & 70.1                          & 4.4          & 25.0          \\
C     & 74.4                          & 3.6          & 2.5           \\ \bottomrule
\end{tabular}
\end{table}

We compare the peak positions in $S(P)$ for all four wafers with respect to the differences in layer thicknesses. We observe a grouping of wafers A and D, which were grown by different methods (CVD and MBE) and characterized by a substantial difference in underlying Si substrate thickness (\SI{725}{\micro m} and \SI{425}{\micro m}). If the Si substrate thickness had a notable impact, this dissimilarity should dominate all other effects. Since this phenomenon is not evident in our observations, we conclude that even the thinner Si substrate serves as an adequate heat sink during the scanning process. This finding holds practical significance since Si substrate thickness is mainly linked to wafer size, a parameter that is typically fixed by the growing tools used and not easily modified.
In terms of similarities, both wafers A and D share the characteristic of having a linearly graded buffer, while wafers B and C have a step-wise graded buffer. Additionally, when considering the combined thickness of the VS (including the GC and CC buffers), wafers A and D exhibit a thickness approximately twice that of wafers B and C. Upon examining the layer structures situated above the virtual substrate, we observe no systematic correlation between power requirements and layer thicknesses among these wafers.
From these observations, we can deduce that the power requirement is predominantly influenced by the specifics of the VS. As SiGe has a worse heat conductivity \cite{Lee1997} compared to Si, the VS therefore acts as a thermal resistor. With increasing thickness, the heat flow to the substrate is reduced, resulting in a higher temperature at the surface and faster regrowth. The grading of the virtual substrate (VS) plays a pivotal role in shaping the quality of the inter-VS interfaces and the distribution of strain within the structure \cite{Paul2004}, which can further modify thermal conductance.
We attribute the further reduction of sheet resistance inside the successful parameter window to an increase in activation ratio with larger temperatures. 

By employing fast optical calibration, we can establish a parameter range for each wafer, ensuring reliable activation. The minimum sheet resistances for each wafer as well as the used laser scan parameters are summarized in Tab. \ref{tab:CalibrationSummary}. The ability to quickly adapt the laser annealing process to different heterostructures is advantageous, especially given the ongoing optimization efforts in Si/SiGe quantum technologies. Our findings reveal that the optimal laser power can vary by nearly a factor of 2 between different wafers (A and B), although their upper layer stacks forming the quantum wells are nearly equivalent. 
With this calibration procedure in place, we can efficiently adapt the laser annealing process to accommodate each new iteration, keeping up with the progress in heterostructure design.

\section{Liquid helium characterization}
\label{sec:Cryo_Chara}
\begin{figure}
\includegraphics{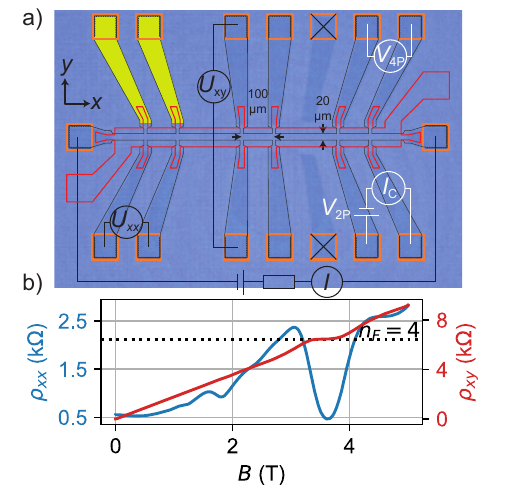}
\caption{\label{fig:hallbar_circuit} Hall-bar measurements: a) Optical microscope image after LA of a sample with implanted contacts (shown in yellow as an example) for a Hall-bar structure. Subsequent fabrication steps are indicated in black (mesa channel etch), orange (contact pad evaporation) and red (gate evaporation). Exemplary circuit diagrams for the 2-point/4-point contact resistance (white) and magneto-transport measurements (black) are shown. The channel of the Hall-bar has a width of \SI{20}{\micro m} and the contact pairs have a spacing of \SI{100}{\micro m} with a spacing of \SI{200}{\micro m} between pairs. Implanted areas with crossed-out contact areas are test structures for fabrication and are not connected to the channel. b) Longitudinal $\rho_{xx}$ and transversal $\rho_{xy}$ resistance obtained by magneto-transport measurements at a gate voltage of $V_\mathrm{G} = \SI{0.6}{V}$. The structure was annealed using LA parameter set S4$^* = $ \PS{1.8}{2.5}. The Hall resistance corresponding to the filling factor $n_\mathrm{F} = 4$ is indicated in the graph.}
\end{figure}

In the next step, we investigate the performance of LA contacts to undoped Si/SiGe quantum wells at \SI{1.6}{K} and \SI{4.2}{K}. We select five parameter sets (S1-S5) from the high success-rate window of wafer A (see Fig. \ref{fig:RT_calibration} (a)) and employ them to fabricate Hall-bar structures. As later applications may be sensitive to the extent of which the thermal input is limited to the contact areas, we also study the lateral resolution of the LA process. For this purpose, we define the region to be annealed so that it ends directly at the junction between the implanted and the gated regions, in which carries are accumulated by a positive gate-voltage. In addition, we utilize each parameter set twice and alternate the orientation of the implanted structure with respect to the scan direction, denoted by $x$ and $y$ in Fig. \ref{fig:hallbar_circuit} (a), respectively. The addition of a star (*) to the parameter name indicates samples where the fast scan axis was oriented along the $y$-direction. The samples used in this process were implanted following the same recipe as the ones used for calibration. In Fig. \ref{fig:hallbar_circuit} (a), an optical microscope image captured after the LA treatment shows a homogeneous color change compared to the previous state observed after implantation. The implanted and subsequently annealed areas are shown in yellow at the example of the upper left pair of contacts. Furthermore, we fabricate reference samples using a global annealing (GA) process at a temperature of \SI{750}{\degree C} for a duration of 30 seconds under an Argon atmosphere. The annealing process is carried out using an \textit{Annealsys AS-One} rapid thermal processing (RTP) system. The subsequent fabrication steps are superimposed on the microscope image in different colors for clarity. The black color represents the mesa etching step, defining the channel region. The orange color signifies the region of metal contact pads that are directly evaporated onto the implanted regions. The red color indicates the gate region, which is isolated from the channel and contacts by a deposited gate oxide layer. Each fabrication step outside of annealing was done parallel on LA and GA samples to ensure comparability. Fabrication details are given in App. \ref{app:sec:fab}.

\subsection{Cryogenic contacts}

To ensure proper connection of the LA contacts down to the 2DEG, we perform magneto-transport measurements at a temperature of \SI{1.6}{K}. An exemplary measurement at a gate voltage of $V_\mathrm{G} = \SI{0.6}{V}$ is shown in Fig. \ref{fig:hallbar_circuit} (b). The structure used was annealed using LA parameter set S4$^* = $ \PS{1.8}{2.5}. We observe a distinct plateau in the longitudinal resistance $\rho_{xy}$ as well as oscillations in the transverse resistance $\rho_{xx}$ which align with the characteristics of the integer quantum Hall effect. The plateau matches the quantized resistance value corresponding to a filling factor of $n_\mathrm{F} = 4$.

The presence of two-dimensional (2D) features proves two important aspects. Firstly, it confirms that the LA contacts retain their conductivity even at cryogenic temperatures, ensuring their suitability for low-temperature application. Secondly, the observation of 2D features verifies that the LA contacts have established contact with the 2DEG.

\subsection{Mobility}
\begin{figure*}
\includegraphics{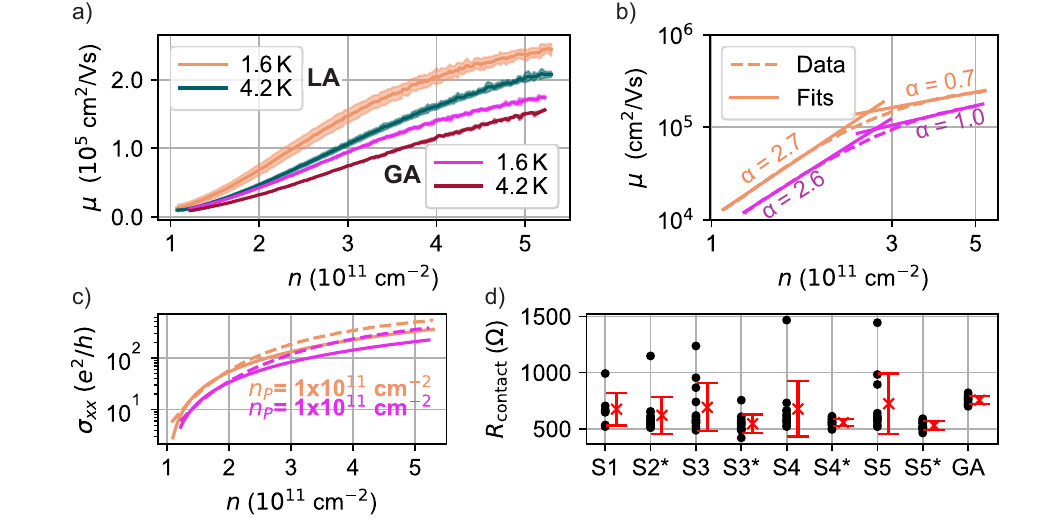}
\caption{\label{fig:cryo_measurements} Cryogenic transport measurements: a) Mobility $\mu$ versus electron density $n$ for the laser annealed (LA) and globally annealed (GA) samples at $\SI{1.6}{K}$ and $\SI{4.2}{K}$. The average of all measured segments is shown as solid lines, while the standard deviation is given as shaded areas. b) Log-log plot of mobility $\mu$ versus electron density $n$. Data sets, corresponding to the color from panel (a), are shown as dashed line while the two fits represent $\mu \propto n^{\alpha}$ for low/high density. Each fit is shown in solid with corresponding power-law exponent $\alpha$. c) Conductivity $\sigma_{xx}$ versus electron density $n$. The data is shown in dashed lines, while the fits $\sigma_{xx} = C (n-n_P)^{1.31}$ to the low electron density regime are shown in solid d) Contact resistance obtained for LA parameter sets as shown in Fig. \ref{fig:RT_calibration}(a) as well as a GA reference sample. Individual data points are shown as black circles while the red crosses indicate average resistance and resistance variance. }
\end{figure*}

Having established the functionality of the LA contacts, in a next step we compare their performance with that of a GA reference sample. To this end, we measure the electron mobility $\mu$ versus electron density $n$ relation for the LA structure annealed with parameter set S4$^*$ as well as for a GA reference at $\SI{1.6}{K}$ and $\SI{4.2}{K}$. The results are shown in Fig. \ref{fig:cryo_measurements} (a). We measure the voltages used to determine the electron density and mobility simultaneously on several pairs of contacts. We combine transverse and longitudinal voltages that share a contact or are measured from neighbouring contacts in order to determine the charge carrier density and corresponding mobility separately for different segments of the channel. When calculating the longitudinal sheet resistance, we take into account the length of the segment. We then calculate the mean value of the electron mobilities at the same electron densities. We take care to stay below a critical gate voltage to prevent the charging of trap states in the heterostructure, oxide, or intervening interfaces, which would screen charged defects resulting in an artificially enhanced mobility \cite{Huang2014,Laroche2015}. We observe that the LA sample has a higher $\mu$ at the same $n$ compared to the GA sample. At $\SI{1.6}{K}$, the LA sample reaches an electron mobility of $\mu = \SI{2.4E5}{cm^2/Vs}$ at an electron density of $n = \SI{5.0E11}{cm^{-2}}$. Meanwhile the GA sample shows a $\mu = \SI{1.7E5}{cm^2/Vs}$ at the same $n$. The electron mobility of both samples decreases as expected when the temperature is increased to $\SI{4.2}{K}$: we obtain $\SI{2.0E5}{cm^2/Vs}$ for the LA sample and $\SI{1.5E5}{cm^2/Vs}$ for the GA sample at the aforementioned density.

To better understand the measurement data, we plot $\mu$ versus $n$ in log-log form and examine the power-law dependence $\mu \propto n^{\alpha}$ as shown in Fig \ref{fig:cryo_measurements} (b). The exponent $\alpha$ depends on the type of the scattering mechanism limiting the mobility. We observe two regions for each sample, following a specific power-law with different exponent $\alpha$. For the LA sample, we get a value of $\alpha = 2.7$ at densities below $n = \SI{2.0E11}{cm^{-2}}$ and a value of $\alpha = 0.7$ at densities above $n = \SI{3.5E11}{cm^{-2}}$. For the GA sample, we obtain $\alpha = 2.6 / 1.0$ in the same regions. This is consistent with previous electron mobility studies conducted in Si/SiGe heterostructures \cite{Laroche2015,Mi2015}. 
In those studies, the low-density scaling is attributed to be limited by the presence of remote charge impurities situated in the vicinity of the semiconductor-oxide interface. Calculations had predicted an exponent of $\alpha_\mathrm{remote} = 3/2$ in such scenarios \cite{Monroe1993,DasSarma2013}. However, it is commonly observed that larger exponents are encountered, which can be explained by incorporating local field corrections \cite{Dolgopolov2003}. As the electron density in the quantum well increases, the influence of these remote impurities decreases, and electron mobility becomes limited by three-dimensional (3D) charge defects within the semiconductor bulk, characterized by an exponent $\alpha_\mathrm{3D} = 1/2$. 

At very low $n$, there is compelling evidence suggesting that the metal-to-insulator transition is primarily governed by disorder-driven electron localization \cite{DasSarma2013}. The presence of an inhomogeneous distribution of charged defects leads to the formation of small and unconnected electron puddles within the system. A conduction path only forms when some of these puddles start connecting. The electron density $n_P$ at this point follows the principles of percolation theory and gives insights about the extent of disorder in the device. To this end, we calculate the longitudinal conductivity $\sigma_{xx}$ as a function of the density and fit the low density regime (up to $\SI{2.0E11}{cm^{-2}}$) according to percolation theory using $\sigma_{xx}(n) = C (n-n_P)^{1.31}$ \cite{Tracy2009}. The results are shown in Fig \ref{fig:cryo_measurements} (c) for $\SI{1.6}{K}$ and in App. \ref{app:sec:4K} for the $\SI{4.2}{K}$ data. We obtain identical $n_P$ for the LA and GA sample at $\SI{1.6}{K}$ of $n_P = \SI{1E11}{cm^{-2}}$.

When we compare the performance of the LA and GA samples, we find that the LA sample exhibits improvements across all transport properties. Nevertheless, it is important to note that a recent comprehensive study assessing electron mobilities and percolation densities in state-of-the-art Si/SiGe heterostructures revealed a sample-to-sample variance on the order of, or even greater than, the differences we observe between the LA and GA samples \cite{PaqueletWuetz2023}. Consequently, it is crucial to gather a more complete statistical dataset to effectively distinguish the impact of the wafer level variability from the potential benefits of laser annealing.

Hence, we have shown that the transport properties of a LA sample compare favorably with those of a GA sample. The similarity in transport characteristics between the two annealing methods suggests that they are interchangeable. Combined with the reduced heat load in the active region, this supports the feasibility of laser annealing as a viable alternative for activation of ion-implanted contacts. This is especially relevant for applications at or below liquid helium temperature, where maintaining the integrity of the active regions is crucial for achieving high-performance quantum devices. 

\subsection{Contact resistance}
In addition to the carrier mobility, we measure and compare the contact resistance at liquid helium temperature (\SI{4.2}{K}) by employing a combination of two- and four-point measurements. Initially, we ramp the gate voltage to $V_\mathrm{G} = \SI{1}{V}$ to achieve a low channel resistance. The exact mobility in the channel is not relevant for this type of measurement. Subsequently, we sweep the voltage $V_\mathrm{2P}$ (see Fig. \ref{fig:hallbar_circuit} (a)) applied across two contacts on one side of the channel and measure the corresponding current $I_\mathrm{C}$ using a lock-in amplifier. Additionally, to obtain the channel resistance, we measure the voltage drop $V_\mathrm{4P}$ across the two contacts on the opposite side of the channel, effectively implementing a four-point measurement configuration. From this we can calculate the sum of the two contact resistances $R_{i,j}$ using
\begin{equation}
    R_{i,j} = R_i + R_j = V_\mathrm{2P} / I_\mathrm{C} - V_\mathrm{4P} / I_\mathrm{C} - R_\mathrm{setup},
\end{equation}
where $R_\mathrm{setup}$ is the series resistance of the setup, which we determine by shorting two measurement lines as close as possible to the sample. By systematically permuting all possible contact combinations $(i,j) \in \{1,2,3,4,5,6\}$ on both sides of the channel, we can establish a system of linear equations and solve it for the single contact resistance $R_{\mathrm{contact},i}$
\begin{equation}
    R_{\mathrm{contact},i} = (R_{i,j} + R_{i,k} - R_{j,k}) / 2. 
\end{equation}

The calculated contact resistances $R_\mathrm{contact}$ for the investigated LA parameter sets (S1-S5) as well as a GA reference sample are shown in Fig. \ref{fig:cryo_measurements} (d). In our investigation, we observe that all laser annealed samples exhibit an average resistance that is lower than that of the reference sample. However, notable variations in resistance values are observed among the LA samples, particularly for samples scanned along the $x$-direction.

We attribute the increased variations observed for samples annealed along the $x$-direction to insufficient overlap between the scanned and the implanted region. This inadequate overlap can lead to a remaining thin amorphous layer between annealed contact and the channel, which imposes a significant series resistance. This can be solved for arbitrary geometries where boundaries of the implanted regions are non-parallel by extending the scanned region beyond the boundaries of the implanted region by one scan-line spacing.

The careful selection of scan orientation relative to the contact geometry, guided by the calibrated LA parameter sets, has allowed us to form contacts at cryogenic temperatures with lower resistivity compared to globally annealed references. These contacts exhibit resistances at least one order of magnitude below the von Klitzing constant $R_\mathrm{R} \approx \SI{26}{\kilo \ohm}$ as a reference for the typical resistance of quantum devices. This emphasizing their suitability for high-performance quantum devices operating at temperatures below \SI{4.2}{K}. This underscores the LA process being a more than suitable alternative to global annealing for cryogenic contact formation.

\section{Conclusion}
Using local laser annealing, we have successfully demonstrated the activation of phosphorus-implanted contacts to a quantum well in undoped Si/SiGe with \SI{530}{\ohm} contact resistance at \SI{4.2}{K}. In the process, we have developed a calibration procedure capable of determining the success of annealing parameters based on optical inspection. We have discovered that the required laser power strongly depends on the details of the virtual SiGe substrate.
This calibration procedure allows us to quickly adapt the annealing parameters to each new heterostructure to keep up with the constant optimization effort. 
When investigating the transport behavior of a laser-annealed Hall-bar structure at a temperature range between \SI{1.6}{K} and \SI{4.2}{K}, we observed that its performance was either comparable or even superior to that of a globally annealed reference sample, with respect to carrier mobility and contact resistance. We obtained an electron mobility of $\mu = \SI{2.4E5}{cm^2/Vs}$ at an electron density of $n = \SI{5.0E11}{cm^{-2}}$ for a laser annealed Hall-bar.
This study demonstrates that laser annealing is a viable alternative for forming ion-implanted contacts in cryogenic applications. The localized treatment of implantation damage provided by laser annealing offers the distinct advantage of minimizing heat impact on critical regions, making it a promising technique to use for quantum devices based on high-performance Si/SiGe heterostructures.

\section*{Acknowledgements}
Implantation of the heterostructures was carried out at Ion Beam Center (IBC) at the Helmholtz-Zentrum Dresden - Rossendorf e. V., a member of the Helmholtz Association. We would like to thank Ulrich Kentsch for assistance. This work was funded by the German Research Foundation (DFG) within the project 289786932 (BO 3140/4-2 and SCHR 1404/2-2).\\

\appendix
\section{Steps along slow axis}
\label{app:sec:steps}
\begin{figure}
\includegraphics{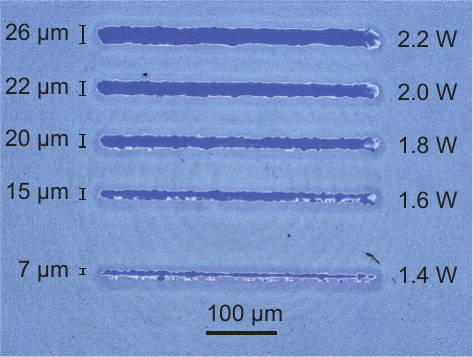}
\caption{\label{fig:Sup_YStep} Optical image of a flat implanted sample of wafer A after individual lines have been annealed with $v = \SI{10}{mm/s}$ and increasing laser power $P$. The colour change indicates recrystallisation, which we use to determine the line width.}
\end{figure}

We reduce the intrinsically three dimensional parameter space \{$P$, $v$, $\Delta y$\} by setting the step size of the slow axis to a fixed value $\Delta y = \SI{10}{\micro m}$. By annealing single lines (here with fixed $v = \SI{10}{mm/s}$ for wafer A), we have determined the width of the recrystallized line depending on the power $P$ as shown in Fig. \ref{fig:Sup_YStep}. Line widths of up to \SI{40}{\micro m} could be achieved ($P = \SI{3.8}{W}$) before signs of surface degradation began to appear. In later quantum devices, it can become very critical to limit the thermal input as much as possible to the contact regions. For this reason, we chose a small step size to maintain high spacial precision: We picked half of the line width of \SI{20}{\micro m}, at which a mostly homogeneous color change along the fast scan direction was observed.
\section{Calibration results B-D}
\label{app:sec:optic}
\begin{figure*}
\includegraphics{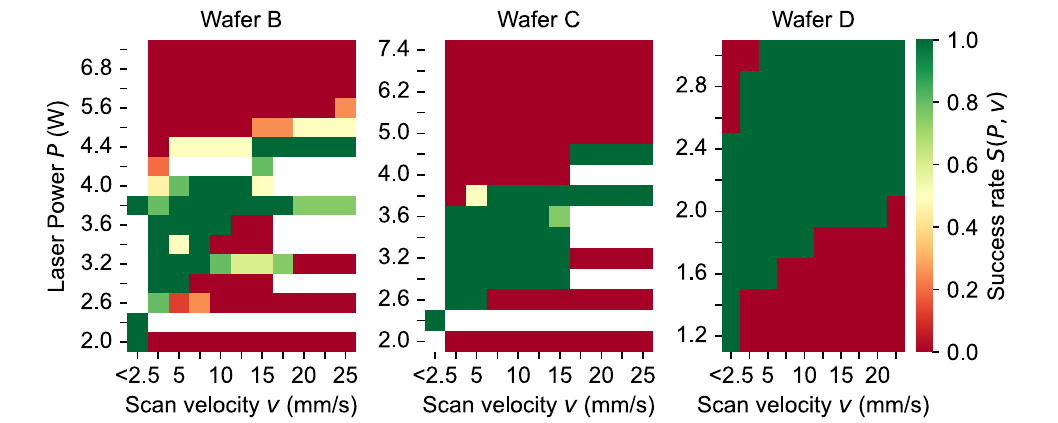}
\caption{\label{fig:Sup_Optic} Map of success rates $S(P,V)$ for wafers B-D obtained by optical inspection of laser annealed squares (cf. Fig. \ref{fig:RT_calibration}a).}
\end{figure*}

The maps of success rate $S$ versus laser annealing parameters for wafers B-D are shown in Fig \ref{fig:Sup_Optic}.
\section{Hall-bar fabrication}
\label{app:sec:fab}
After being diced into \SI{10}{\milli \meter}x\SI{10}{\milli \meter} pieces, the samples underwent a cleaning process using an RCA flow. This cleaning procedure involved consecutive room temperature baths in Piranha (H$_2$SO$_4$:H$_2$O$_2$ 3:1), SC1 (NH$_3$:H$_2$O$_2$:H$_2$O 1:1:5), and SC2 (HCl:H$_2$O$_2$:H$_2$O 1:1:6) solutions, with interleaved hydrofluoric acid (HF) oxide strips and DI-water rinsing steps. Afterwards, markers for layer-on-layer alignment were etched into the surface using reactive ion etching (RIE). Optical photo resist was used as mask to define the contact areas. Implantation was carried out by the Institute of Ion Beam Physics and Materials (Helmholtz-Zentrum-Dresden-Rossendorf HZDR) in a \textit{Danfysik A/S, Denmark, Model 1050}. Following the implantation, the resist mask was subsequently removed using a solvent, and an additional RCA cleaning step was performed. The samples were then subjected to either laser or global annealing, which was carried in the mentioned RTP system at $\SI{750}{^\circ}$C for \SI{30}{s} in an argon atmosphere. After annealing, the active channel region was defined using a RIE etching process. To remove any possible native oxide, an HF dip was performed prior to the evaporation of the metal contact pads. The active region was isolated from the subsequently evaporated gate by a \SI{20}{nm} thick layer of aluminum oxide, deposited via atomic layer deposition.

\section{\SI{4}{K} measurement results}
 \label{app:sec:4K}
\begin{figure}
\includegraphics{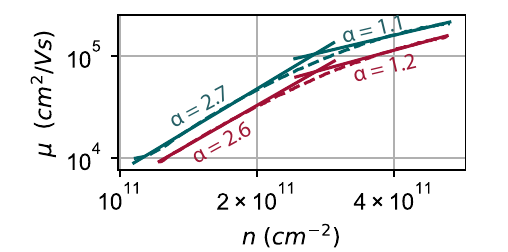}
\caption{\label{fig:Sup_4K_alpha} Log-log plot of electron mobility $\mu$ versus electron density $n$ for LA and GA sample, measured at \SI{4}{K}. Dashed lines indicate data, while filled lines show two-part wise power law fits. Colors correspond to the legend in Fig. \ref{fig:cryo_measurements}.}
\end{figure}

\begin{figure}
\includegraphics{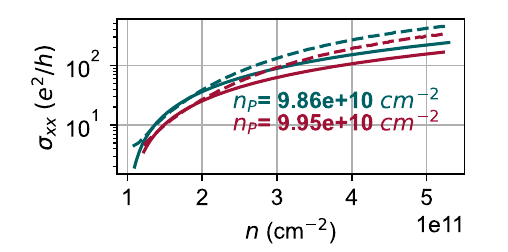}
\caption{\label{fig:Sup_4K_perc} Longitudinal conductance $\sigma_{xx}$ versus carrier density $n$ for LA and GA sample, measured at \SI{4}{K}. Solid lines indicated fit functions consistent with percolation theory.}
\end{figure}

The results of the power law $\mu \propto n^{\alpha}$ fit, obtained for the LA and GA sample at \SI{4}{K} are shown in Fig. \ref{fig:Sup_4K_alpha}. The colors used correspond to the legend in Fig. \ref{fig:cryo_measurements}. Furthermore, the longitudinal conductance $\sigma_{xx}$ versus mobility $\mu$ plot for these samples, as well as the extracted percolation density, are given in Fig. \ref{fig:Sup_4K_perc}.

\begin{thebibliography}{49}%
\makeatletter
\providecommand \@ifxundefined [1]{%
 \@ifx{#1\undefined}
}%
\providecommand \@ifnum [1]{%
 \ifnum #1\expandafter \@firstoftwo
 \else \expandafter \@secondoftwo
 \fi
}%
\providecommand \@ifx [1]{%
 \ifx #1\expandafter \@firstoftwo
 \else \expandafter \@secondoftwo
 \fi
}%
\providecommand \natexlab [1]{#1}%
\providecommand \enquote  [1]{``#1''}%
\providecommand \bibnamefont  [1]{#1}%
\providecommand \bibfnamefont [1]{#1}%
\providecommand \citenamefont [1]{#1}%
\providecommand \href@noop [0]{\@secondoftwo}%
\providecommand \href [0]{\begingroup \@sanitize@url \@href}%
\providecommand \@href[1]{\@@startlink{#1}\@@href}%
\providecommand \@@href[1]{\endgroup#1\@@endlink}%
\providecommand \@sanitize@url [0]{\catcode `\\12\catcode `\$12\catcode
  `\&12\catcode `\#12\catcode `\^12\catcode `\_12\catcode `\%12\relax}%
\providecommand \@@startlink[1]{}%
\providecommand \@@endlink[0]{}%
\providecommand \url  [0]{\begingroup\@sanitize@url \@url }%
\providecommand \@url [1]{\endgroup\@href {#1}{\urlprefix }}%
\providecommand \urlprefix  [0]{URL }%
\providecommand \Eprint [0]{\href }%
\providecommand \doibase [0]{https://doi.org/}%
\providecommand \selectlanguage [0]{\@gobble}%
\providecommand \bibinfo  [0]{\@secondoftwo}%
\providecommand \bibfield  [0]{\@secondoftwo}%
\providecommand \translation [1]{[#1]}%
\providecommand \BibitemOpen [0]{}%
\providecommand \bibitemStop [0]{}%
\providecommand \bibitemNoStop [0]{.\EOS\space}%
\providecommand \EOS [0]{\spacefactor3000\relax}%
\providecommand \BibitemShut  [1]{\csname bibitem#1\endcsname}%
\let\auto@bib@innerbib\@empty
\bibitem [{\citenamefont {Nazir}\ \emph {et~al.}(2022)\citenamefont {Nazir},
  \citenamefont {Kushwaha}, \citenamefont {Pampori}, \citenamefont {Ahsan},\
  and\ \citenamefont {Chauhan}}]{Nazir2022}%
  \BibitemOpen
  \bibfield  {author} {\bibinfo {author} {\bibfnamefont {M.~S.}\ \bibnamefont
  {Nazir}}, \bibinfo {author} {\bibfnamefont {P.}~\bibnamefont {Kushwaha}},
  \bibinfo {author} {\bibfnamefont {A.}~\bibnamefont {Pampori}}, \bibinfo
  {author} {\bibfnamefont {S.~A.}\ \bibnamefont {Ahsan}},\ and\ \bibinfo
  {author} {\bibfnamefont {Y.~S.}\ \bibnamefont {Chauhan}},\ }\href
  {https://doi.org/10.1109/TED.2022.3204523} {\bibfield  {journal} {\bibinfo
  {journal} {IEEE Trans. Electron Devices}\ }\textbf {\bibinfo {volume} {69}},\
  \bibinfo {pages} {6016} (\bibinfo {year} {2022})}\BibitemShut {NoStop}%
\bibitem [{\citenamefont {Zeng}\ \emph {et~al.}(2022)\citenamefont {Zeng},
  \citenamefont {Zhang}, \citenamefont {Luo}, \citenamefont {Xiang},
  \citenamefont {Zhang}, \citenamefont {Wen}, \citenamefont {Xue},
  \citenamefont {Hu}, \citenamefont {Sun}, \citenamefont {Yang}, \citenamefont
  {Sun},\ and\ \citenamefont {Guo}}]{Zeng2022}%
  \BibitemOpen
  \bibfield  {author} {\bibinfo {author} {\bibfnamefont {B.}~\bibnamefont
  {Zeng}}, \bibinfo {author} {\bibfnamefont {H.}~\bibnamefont {Zhang}},
  \bibinfo {author} {\bibfnamefont {C.}~\bibnamefont {Luo}}, \bibinfo {author}
  {\bibfnamefont {Z.}~\bibnamefont {Xiang}}, \bibinfo {author} {\bibfnamefont
  {Y.}~\bibnamefont {Zhang}}, \bibinfo {author} {\bibfnamefont
  {M.}~\bibnamefont {Wen}}, \bibinfo {author} {\bibfnamefont {Q.}~\bibnamefont
  {Xue}}, \bibinfo {author} {\bibfnamefont {S.}~\bibnamefont {Hu}}, \bibinfo
  {author} {\bibfnamefont {Y.}~\bibnamefont {Sun}}, \bibinfo {author}
  {\bibfnamefont {L.}~\bibnamefont {Yang}}, \bibinfo {author} {\bibfnamefont
  {H.}~\bibnamefont {Sun}},\ and\ \bibinfo {author} {\bibfnamefont
  {G.}~\bibnamefont {Guo}},\ }\href
  {https://iopscience.iop.org/article/10.1088/1361-6463/ac89fc/meta} {\bibfield
   {journal} {\bibinfo  {journal} {J. Phys. D.}\ }\textbf {\bibinfo {volume}
  {55}},\ \bibinfo {pages} {434003} (\bibinfo {year} {2022})}\BibitemShut
  {NoStop}%
\bibitem [{\citenamefont {Najafizadeh}\ \emph {et~al.}(2006)\citenamefont
  {Najafizadeh}, \citenamefont {Zhu}, \citenamefont {Krithivasan},
  \citenamefont {Cressler}, \citenamefont {Cui}, \citenamefont {Niu},
  \citenamefont {Chen}, \citenamefont {Ulaganathan}, \citenamefont {Blalock},\
  and\ \citenamefont {Joseph}}]{Najafizadeh2006}%
  \BibitemOpen
  \bibfield  {author} {\bibinfo {author} {\bibfnamefont {L.}~\bibnamefont
  {Najafizadeh}}, \bibinfo {author} {\bibfnamefont {C.}~\bibnamefont {Zhu}},
  \bibinfo {author} {\bibfnamefont {R.}~\bibnamefont {Krithivasan}}, \bibinfo
  {author} {\bibfnamefont {J.~D.}\ \bibnamefont {Cressler}}, \bibinfo {author}
  {\bibfnamefont {Y.}~\bibnamefont {Cui}}, \bibinfo {author} {\bibfnamefont
  {G.}~\bibnamefont {Niu}}, \bibinfo {author} {\bibfnamefont {S.}~\bibnamefont
  {Chen}}, \bibinfo {author} {\bibfnamefont {C.}~\bibnamefont {Ulaganathan}},
  \bibinfo {author} {\bibfnamefont {B.~J.}\ \bibnamefont {Blalock}},\ and\
  \bibinfo {author} {\bibfnamefont {A.~J.}\ \bibnamefont {Joseph}},\ }\bibfield
   {journal} {\bibinfo  {journal} {Proc. IEEE}\ }\href
  {https://doi.org/10.1109/BIPOL.2006.311117} {10.1109/BIPOL.2006.311117}
  (\bibinfo {year} {2006})\BibitemShut {NoStop}%
\bibitem [{\citenamefont {Ward}\ \emph {et~al.}(2006)\citenamefont {Ward},
  \citenamefont {Dawson}, \citenamefont {Zhu}, \citenamefont {Kirschman},
  \citenamefont {Niu}, \citenamefont {Nelms}, \citenamefont {Mueller},
  \citenamefont {Hennessy}, \citenamefont {Mueller}, \citenamefont {Patterson},
  \citenamefont {Dickman},\ and\ \citenamefont {Hammoud}}]{Ward2006}%
  \BibitemOpen
  \bibfield  {author} {\bibinfo {author} {\bibfnamefont {R.~R.}\ \bibnamefont
  {Ward}}, \bibinfo {author} {\bibfnamefont {W.~J.}\ \bibnamefont {Dawson}},
  \bibinfo {author} {\bibfnamefont {L.}~\bibnamefont {Zhu}}, \bibinfo {author}
  {\bibfnamefont {R.~K.}\ \bibnamefont {Kirschman}}, \bibinfo {author}
  {\bibfnamefont {G.}~\bibnamefont {Niu}}, \bibinfo {author} {\bibfnamefont
  {R.~M.}\ \bibnamefont {Nelms}}, \bibinfo {author} {\bibfnamefont
  {O.}~\bibnamefont {Mueller}}, \bibinfo {author} {\bibfnamefont {M.~J.}\
  \bibnamefont {Hennessy}}, \bibinfo {author} {\bibfnamefont {E.~K.}\
  \bibnamefont {Mueller}}, \bibinfo {author} {\bibfnamefont {R.~L.}\
  \bibnamefont {Patterson}}, \bibinfo {author} {\bibfnamefont {J.~E.}\
  \bibnamefont {Dickman}},\ and\ \bibinfo {author} {\bibfnamefont
  {A.}~\bibnamefont {Hammoud}},\ }\href
  {https://doi.org/10.1109/apec.2006.1620766} {\bibfield  {journal} {\bibinfo
  {journal} {Proc. IEEE}\ }\textbf {\bibinfo {volume} {2006}},\ \bibinfo
  {pages} {1673} (\bibinfo {year} {2006})}\BibitemShut {NoStop}%
\bibitem [{\citenamefont {Hollmann}\ \emph {et~al.}(2018)\citenamefont
  {Hollmann}, \citenamefont {Jirovec}, \citenamefont {Kucharski}, \citenamefont
  {Kissinger}, \citenamefont {Fischer},\ and\ \citenamefont
  {Schreiber}}]{Hollmann2018}%
  \BibitemOpen
  \bibfield  {author} {\bibinfo {author} {\bibfnamefont {A.}~\bibnamefont
  {Hollmann}}, \bibinfo {author} {\bibfnamefont {D.}~\bibnamefont {Jirovec}},
  \bibinfo {author} {\bibfnamefont {M.}~\bibnamefont {Kucharski}}, \bibinfo
  {author} {\bibfnamefont {D.}~\bibnamefont {Kissinger}}, \bibinfo {author}
  {\bibfnamefont {G.}~\bibnamefont {Fischer}},\ and\ \bibinfo {author}
  {\bibfnamefont {L.~R.}\ \bibnamefont {Schreiber}},\ }\href
  {https://doi.org/10.1063/1.5038258} {\bibfield  {journal} {\bibinfo
  {journal} {Rev. Sci. Instrum.}\ }\textbf {\bibinfo {volume} {89}},\ \bibinfo
  {pages} {114701} (\bibinfo {year} {2018})}\BibitemShut {NoStop}%
\bibitem [{\citenamefont {Hosseini}\ and\ \citenamefont
  {Bardin}(2021)}]{Hosseini2021}%
  \BibitemOpen
  \bibfield  {author} {\bibinfo {author} {\bibfnamefont {M.}~\bibnamefont
  {Hosseini}}\ and\ \bibinfo {author} {\bibfnamefont {J.~C.}\ \bibnamefont
  {Bardin}},\ }\href {https://doi.org/10.1109/IMS19712.2021.9574993} {\bibfield
   {journal} {\bibinfo  {journal} {IEEE MTT-S IMS}\ ,\ \bibinfo {pages} {896}}
  (\bibinfo {year} {2021})}\BibitemShut {NoStop}%
\bibitem [{\citenamefont {Yoneda}\ \emph {et~al.}(2018)\citenamefont {Yoneda},
  \citenamefont {Takeda}, \citenamefont {Otsuka}, \citenamefont {Nakajima},
  \citenamefont {Delbecq}, \citenamefont {Allison}, \citenamefont {Honda},
  \citenamefont {Kodera}, \citenamefont {Oda}, \citenamefont {Hoshi},
  \citenamefont {Usami}, \citenamefont {Itoh},\ and\ \citenamefont
  {Tarucha}}]{Yoneda2018}%
  \BibitemOpen
  \bibfield  {author} {\bibinfo {author} {\bibfnamefont {J.}~\bibnamefont
  {Yoneda}}, \bibinfo {author} {\bibfnamefont {K.}~\bibnamefont {Takeda}},
  \bibinfo {author} {\bibfnamefont {T.}~\bibnamefont {Otsuka}}, \bibinfo
  {author} {\bibfnamefont {T.}~\bibnamefont {Nakajima}}, \bibinfo {author}
  {\bibfnamefont {M.~R.}\ \bibnamefont {Delbecq}}, \bibinfo {author}
  {\bibfnamefont {G.}~\bibnamefont {Allison}}, \bibinfo {author} {\bibfnamefont
  {T.}~\bibnamefont {Honda}}, \bibinfo {author} {\bibfnamefont
  {T.}~\bibnamefont {Kodera}}, \bibinfo {author} {\bibfnamefont
  {S.}~\bibnamefont {Oda}}, \bibinfo {author} {\bibfnamefont {Y.}~\bibnamefont
  {Hoshi}}, \bibinfo {author} {\bibfnamefont {N.}~\bibnamefont {Usami}},
  \bibinfo {author} {\bibfnamefont {K.~M.}\ \bibnamefont {Itoh}},\ and\
  \bibinfo {author} {\bibfnamefont {S.}~\bibnamefont {Tarucha}},\ }\href
  {https://doi.org/10.1038/s41565-017-0014-x} {\bibfield  {journal} {\bibinfo
  {journal} {Nat. Nanotechnol.}\ }\textbf {\bibinfo {volume} {13}},\ \bibinfo
  {pages} {102} (\bibinfo {year} {2018})}\BibitemShut {NoStop}%
\bibitem [{\citenamefont {Noiri}\ \emph {et~al.}(2022)\citenamefont {Noiri},
  \citenamefont {Takeda}, \citenamefont {Nakajima}, \citenamefont {Kobayashi},
  \citenamefont {Sammak}, \citenamefont {Scappucci},\ and\ \citenamefont
  {Tarucha}}]{Noiri2022}%
  \BibitemOpen
  \bibfield  {author} {\bibinfo {author} {\bibfnamefont {A.}~\bibnamefont
  {Noiri}}, \bibinfo {author} {\bibfnamefont {K.}~\bibnamefont {Takeda}},
  \bibinfo {author} {\bibfnamefont {T.}~\bibnamefont {Nakajima}}, \bibinfo
  {author} {\bibfnamefont {T.}~\bibnamefont {Kobayashi}}, \bibinfo {author}
  {\bibfnamefont {A.}~\bibnamefont {Sammak}}, \bibinfo {author} {\bibfnamefont
  {G.}~\bibnamefont {Scappucci}},\ and\ \bibinfo {author} {\bibfnamefont
  {S.}~\bibnamefont {Tarucha}},\ }\href
  {https://doi.org/10.1038/s41586-021-04182-y} {\bibfield  {journal} {\bibinfo
  {journal} {Nature}\ }\textbf {\bibinfo {volume} {601}},\ \bibinfo {pages}
  {338} (\bibinfo {year} {2022})}\BibitemShut {NoStop}%
\bibitem [{\citenamefont {Xue}\ \emph {et~al.}(2022)\citenamefont {Xue},
  \citenamefont {Russ}, \citenamefont {Samkharadze}, \citenamefont {Undseth},
  \citenamefont {Sammak}, \citenamefont {Scappucci},\ and\ \citenamefont
  {Vandersypen}}]{Xue2022}%
  \BibitemOpen
  \bibfield  {author} {\bibinfo {author} {\bibfnamefont {X.}~\bibnamefont
  {Xue}}, \bibinfo {author} {\bibfnamefont {M.}~\bibnamefont {Russ}}, \bibinfo
  {author} {\bibfnamefont {N.}~\bibnamefont {Samkharadze}}, \bibinfo {author}
  {\bibfnamefont {B.}~\bibnamefont {Undseth}}, \bibinfo {author} {\bibfnamefont
  {A.}~\bibnamefont {Sammak}}, \bibinfo {author} {\bibfnamefont
  {G.}~\bibnamefont {Scappucci}},\ and\ \bibinfo {author} {\bibfnamefont
  {L.~M.}\ \bibnamefont {Vandersypen}},\ }\href
  {https://doi.org/10.1038/s41586-021-04273-w} {\bibfield  {journal} {\bibinfo
  {journal} {Nature}\ }\textbf {\bibinfo {volume} {601}},\ \bibinfo {pages}
  {343} (\bibinfo {year} {2022})}\BibitemShut {NoStop}%
\bibitem [{\citenamefont {Struck}\ \emph {et~al.}(2023)\citenamefont {Struck},
  \citenamefont {Volmer}, \citenamefont {Visser}, \citenamefont {Offermann},
  \citenamefont {Xue}, \citenamefont {Tu}, \citenamefont {Trellenkamp},
  \citenamefont {Cywi{\'{n}}ski}, \citenamefont {Bluhm},\ and\ \citenamefont
  {Schreiber}}]{Struck2023}%
  \BibitemOpen
  \bibfield  {author} {\bibinfo {author} {\bibfnamefont {T.}~\bibnamefont
  {Struck}}, \bibinfo {author} {\bibfnamefont {M.}~\bibnamefont {Volmer}},
  \bibinfo {author} {\bibfnamefont {L.}~\bibnamefont {Visser}}, \bibinfo
  {author} {\bibfnamefont {T.}~\bibnamefont {Offermann}}, \bibinfo {author}
  {\bibfnamefont {R.}~\bibnamefont {Xue}}, \bibinfo {author} {\bibfnamefont
  {J.-S.}\ \bibnamefont {Tu}}, \bibinfo {author} {\bibfnamefont
  {S.}~\bibnamefont {Trellenkamp}}, \bibinfo {author} {\bibfnamefont
  {{\L}.}~\bibnamefont {Cywi{\'{n}}ski}}, \bibinfo {author} {\bibfnamefont
  {H.}~\bibnamefont {Bluhm}},\ and\ \bibinfo {author} {\bibfnamefont {L.~R.}\
  \bibnamefont {Schreiber}},\ }\href {http://arxiv.org/abs/2307.04897}
  {\bibfield  {journal} {\bibinfo  {journal} {arxiv: 2307.04897}\ } (\bibinfo
  {year} {2023})}\BibitemShut {NoStop}%
\bibitem [{\citenamefont {Xue}\ \emph {et~al.}(2023)\citenamefont {Xue},
  \citenamefont {Beer}, \citenamefont {Seidler}, \citenamefont {Humpohl},
  \citenamefont {Tu}, \citenamefont {Trellenkamp}, \citenamefont {Struck},
  \citenamefont {Bluhm},\ and\ \citenamefont {Schreiber}}]{Xue2023}%
  \BibitemOpen
  \bibfield  {author} {\bibinfo {author} {\bibfnamefont {R.}~\bibnamefont
  {Xue}}, \bibinfo {author} {\bibfnamefont {M.}~\bibnamefont {Beer}}, \bibinfo
  {author} {\bibfnamefont {I.}~\bibnamefont {Seidler}}, \bibinfo {author}
  {\bibfnamefont {S.}~\bibnamefont {Humpohl}}, \bibinfo {author} {\bibfnamefont
  {J.-S.}\ \bibnamefont {Tu}}, \bibinfo {author} {\bibfnamefont
  {S.}~\bibnamefont {Trellenkamp}}, \bibinfo {author} {\bibfnamefont
  {T.}~\bibnamefont {Struck}}, \bibinfo {author} {\bibfnamefont
  {H.}~\bibnamefont {Bluhm}},\ and\ \bibinfo {author} {\bibfnamefont {L.~R.}\
  \bibnamefont {Schreiber}},\ }\href {http://arxiv.org/abs/2306.16375}
  {\bibfield  {journal} {\bibinfo  {journal} {arxiv: 2306.16375}\ } (\bibinfo
  {year} {2023})}\BibitemShut {NoStop}%
\bibitem [{\citenamefont {Maune}\ \emph {et~al.}(2012)\citenamefont {Maune},
  \citenamefont {Borselli}, \citenamefont {Huang}, \citenamefont {Ladd},
  \citenamefont {Deelman}, \citenamefont {Holabird}, \citenamefont {Kiselev},
  \citenamefont {Alvarado-Rodriguez}, \citenamefont {Ross}, \citenamefont
  {Schmitz}, \citenamefont {Sokolich}, \citenamefont {Watson}, \citenamefont
  {Gyure},\ and\ \citenamefont {Hunter}}]{Maune2012}%
  \BibitemOpen
  \bibfield  {author} {\bibinfo {author} {\bibfnamefont {B.~M.}\ \bibnamefont
  {Maune}}, \bibinfo {author} {\bibfnamefont {M.~G.}\ \bibnamefont {Borselli}},
  \bibinfo {author} {\bibfnamefont {B.}~\bibnamefont {Huang}}, \bibinfo
  {author} {\bibfnamefont {T.~D.}\ \bibnamefont {Ladd}}, \bibinfo {author}
  {\bibfnamefont {P.~W.}\ \bibnamefont {Deelman}}, \bibinfo {author}
  {\bibfnamefont {K.~S.}\ \bibnamefont {Holabird}}, \bibinfo {author}
  {\bibfnamefont {A.~A.}\ \bibnamefont {Kiselev}}, \bibinfo {author}
  {\bibfnamefont {I.}~\bibnamefont {Alvarado-Rodriguez}}, \bibinfo {author}
  {\bibfnamefont {R.~S.}\ \bibnamefont {Ross}}, \bibinfo {author}
  {\bibfnamefont {A.~E.}\ \bibnamefont {Schmitz}}, \bibinfo {author}
  {\bibfnamefont {M.}~\bibnamefont {Sokolich}}, \bibinfo {author}
  {\bibfnamefont {C.~A.}\ \bibnamefont {Watson}}, \bibinfo {author}
  {\bibfnamefont {M.~F.}\ \bibnamefont {Gyure}},\ and\ \bibinfo {author}
  {\bibfnamefont {A.~T.}\ \bibnamefont {Hunter}},\ }\href
  {https://doi.org/10.1038/nature10707} {\bibfield  {journal} {\bibinfo
  {journal} {Nature}\ }\textbf {\bibinfo {volume} {481}},\ \bibinfo {pages}
  {344} (\bibinfo {year} {2012})}\BibitemShut {NoStop}%
\bibitem [{\citenamefont {Klos~et al.}(2023)}]{Bougeard}%
  \BibitemOpen
  \bibfield  {author} {\bibinfo {author} {\bibfnamefont {J.}~\bibnamefont
  {Klos~et al.}},\ }\href@noop {} {\bibinfo {title} {{(In Preperation)}}}
  (\bibinfo {year} {2023})\BibitemShut {NoStop}%
\bibitem [{\citenamefont {Liu}\ \emph {et~al.}(2023)\citenamefont {Liu},
  \citenamefont {Gradwohl}, \citenamefont {Lu}, \citenamefont {Dadyis},
  \citenamefont {Yamamoto}, \citenamefont {Becker}, \citenamefont {Storck},
  \citenamefont {Remmele}, \citenamefont {Boeck}, \citenamefont {Richter},\
  and\ \citenamefont {Albrecht}}]{Liu2022b}%
  \BibitemOpen
  \bibfield  {author} {\bibinfo {author} {\bibfnamefont {Y.}~\bibnamefont
  {Liu}}, \bibinfo {author} {\bibfnamefont {K.-p.}\ \bibnamefont {Gradwohl}},
  \bibinfo {author} {\bibfnamefont {C.-h.}\ \bibnamefont {Lu}}, \bibinfo
  {author} {\bibfnamefont {K.}~\bibnamefont {Dadyis}}, \bibinfo {author}
  {\bibfnamefont {Y.}~\bibnamefont {Yamamoto}}, \bibinfo {author}
  {\bibfnamefont {L.}~\bibnamefont {Becker}}, \bibinfo {author} {\bibfnamefont
  {P.}~\bibnamefont {Storck}}, \bibinfo {author} {\bibfnamefont
  {T.}~\bibnamefont {Remmele}}, \bibinfo {author} {\bibfnamefont
  {T.}~\bibnamefont {Boeck}}, \bibinfo {author} {\bibfnamefont
  {C.}~\bibnamefont {Richter}},\ and\ \bibinfo {author} {\bibfnamefont
  {M.}~\bibnamefont {Albrecht}},\ }\href {https://doi.org/10.1063/5.0155448}
  {\bibfield  {journal} {\bibinfo  {journal} {J. Appl. Phys.}\ }\textbf
  {\bibinfo {volume} {134}},\ \bibinfo {pages} {035302} (\bibinfo {year}
  {2023})}\BibitemShut {NoStop}%
\bibitem [{\citenamefont {Losert}\ \emph {et~al.}(2023)\citenamefont {Losert},
  \citenamefont {Eriksson}, \citenamefont {Joynt}, \citenamefont {Rahman},
  \citenamefont {Scappucci}, \citenamefont {Coppersmith},\ and\ \citenamefont
  {Friesen}}]{Losert2023}%
  \BibitemOpen
  \bibfield  {author} {\bibinfo {author} {\bibfnamefont {M.~P.}\ \bibnamefont
  {Losert}}, \bibinfo {author} {\bibfnamefont {M.~A.}\ \bibnamefont
  {Eriksson}}, \bibinfo {author} {\bibfnamefont {R.}~\bibnamefont {Joynt}},
  \bibinfo {author} {\bibfnamefont {R.}~\bibnamefont {Rahman}}, \bibinfo
  {author} {\bibfnamefont {G.}~\bibnamefont {Scappucci}}, \bibinfo {author}
  {\bibfnamefont {S.~N.}\ \bibnamefont {Coppersmith}},\ and\ \bibinfo {author}
  {\bibfnamefont {M.}~\bibnamefont {Friesen}},\ }\href
  {http://arxiv.org/abs/2303.02499} {\bibfield  {journal} {\bibinfo  {journal}
  {arxiv: 2303.02499}\ } (\bibinfo {year} {2023})}\BibitemShut {NoStop}%
\bibitem [{\citenamefont {Liu}\ \emph {et~al.}(2022)\citenamefont {Liu},
  \citenamefont {Gradwohl}, \citenamefont {Lu}, \citenamefont {Remmele},
  \citenamefont {Yamamoto}, \citenamefont {Zoellner}, \citenamefont
  {Schroeder}, \citenamefont {Boeck}, \citenamefont {Amari}, \citenamefont
  {Richter},\ and\ \citenamefont {Albrecht}}]{Liu2022}%
  \BibitemOpen
  \bibfield  {author} {\bibinfo {author} {\bibfnamefont {Y.}~\bibnamefont
  {Liu}}, \bibinfo {author} {\bibfnamefont {K.~P.}\ \bibnamefont {Gradwohl}},
  \bibinfo {author} {\bibfnamefont {C.~H.}\ \bibnamefont {Lu}}, \bibinfo
  {author} {\bibfnamefont {T.}~\bibnamefont {Remmele}}, \bibinfo {author}
  {\bibfnamefont {Y.}~\bibnamefont {Yamamoto}}, \bibinfo {author}
  {\bibfnamefont {M.~H.}\ \bibnamefont {Zoellner}}, \bibinfo {author}
  {\bibfnamefont {T.}~\bibnamefont {Schroeder}}, \bibinfo {author}
  {\bibfnamefont {T.}~\bibnamefont {Boeck}}, \bibinfo {author} {\bibfnamefont
  {H.}~\bibnamefont {Amari}}, \bibinfo {author} {\bibfnamefont
  {C.}~\bibnamefont {Richter}},\ and\ \bibinfo {author} {\bibfnamefont
  {M.}~\bibnamefont {Albrecht}},\ }\href {https://doi.org/10.1063/5.0101753}
  {\bibfield  {journal} {\bibinfo  {journal} {J. Appl. Phys.}\ }\textbf
  {\bibinfo {volume} {132}},\ \bibinfo {pages} {085302} (\bibinfo {year}
  {2022})}\BibitemShut {NoStop}%
\bibitem [{\citenamefont {McJunkin}\ \emph {et~al.}(2021)\citenamefont
  {McJunkin}, \citenamefont {Macquarrie}, \citenamefont {Tom}, \citenamefont
  {Neyens}, \citenamefont {Dodson}, \citenamefont {Thorgrimsson}, \citenamefont
  {Corrigan}, \citenamefont {Ercan}, \citenamefont {Savage}, \citenamefont
  {Lagally}, \citenamefont {Joynt}, \citenamefont {Coppersmith}, \citenamefont
  {Friesen},\ and\ \citenamefont {Eriksson}}]{McJunkin2021}%
  \BibitemOpen
  \bibfield  {author} {\bibinfo {author} {\bibfnamefont {T.}~\bibnamefont
  {McJunkin}}, \bibinfo {author} {\bibfnamefont {E.~R.}\ \bibnamefont
  {MacQuarrie}}, \bibinfo {author} {\bibfnamefont {L.}~\bibnamefont {Tom}},
  \bibinfo {author} {\bibfnamefont {S.~F.}\ \bibnamefont {Neyens}}, \bibinfo
  {author} {\bibfnamefont {J.~P.}\ \bibnamefont {Dodson}}, \bibinfo {author}
  {\bibfnamefont {B.}~\bibnamefont {Thorgrimsson}}, \bibinfo {author}
  {\bibfnamefont {J.}~\bibnamefont {Corrigan}}, \bibinfo {author}
  {\bibfnamefont {H.~E.}\ \bibnamefont {Ercan}}, \bibinfo {author}
  {\bibfnamefont {D.~E.}\ \bibnamefont {Savage}}, \bibinfo {author}
  {\bibfnamefont {M.~G.}\ \bibnamefont {Lagally}}, \bibinfo {author}
  {\bibfnamefont {R.}~\bibnamefont {Joynt}}, \bibinfo {author} {\bibfnamefont
  {S.~N.}\ \bibnamefont {Coppersmith}}, \bibinfo {author} {\bibfnamefont
  {M.}~\bibnamefont {Friesen}},\ and\ \bibinfo {author} {\bibfnamefont {M.~A.}\
  \bibnamefont {Eriksson}},\ }\href
  {https://doi.org/10.1103/PhysRevB.104.085406} {\bibfield  {journal} {\bibinfo
   {journal} {Phys. Rev. B}\ }\textbf {\bibinfo {volume} {104}},\ \bibinfo
  {pages} {085406} (\bibinfo {year} {2021})}\BibitemShut {NoStop}%
\bibitem [{\citenamefont {McJunkin}\ \emph {et~al.}(2022)\citenamefont
  {McJunkin}, \citenamefont {Harpt}, \citenamefont {Feng}, \citenamefont
  {Losert}, \citenamefont {Rahman}, \citenamefont {Dodson}, \citenamefont
  {Wolfe}, \citenamefont {Savage}, \citenamefont {Lagally}, \citenamefont
  {Coppersmith}, \citenamefont {Friesen}, \citenamefont {Joynt},\ and\
  \citenamefont {Eriksson}}]{McJunkin2022}%
  \BibitemOpen
  \bibfield  {author} {\bibinfo {author} {\bibfnamefont {T.}~\bibnamefont
  {McJunkin}}, \bibinfo {author} {\bibfnamefont {B.}~\bibnamefont {Harpt}},
  \bibinfo {author} {\bibfnamefont {Y.}~\bibnamefont {Feng}}, \bibinfo {author}
  {\bibfnamefont {M.~P.}\ \bibnamefont {Losert}}, \bibinfo {author}
  {\bibfnamefont {R.}~\bibnamefont {Rahman}}, \bibinfo {author} {\bibfnamefont
  {J.~P.}\ \bibnamefont {Dodson}}, \bibinfo {author} {\bibfnamefont {M.~A.}\
  \bibnamefont {Wolfe}}, \bibinfo {author} {\bibfnamefont {D.~E.}\ \bibnamefont
  {Savage}}, \bibinfo {author} {\bibfnamefont {M.~G.}\ \bibnamefont {Lagally}},
  \bibinfo {author} {\bibfnamefont {S.~N.}\ \bibnamefont {Coppersmith}},
  \bibinfo {author} {\bibfnamefont {M.}~\bibnamefont {Friesen}}, \bibinfo
  {author} {\bibfnamefont {R.}~\bibnamefont {Joynt}},\ and\ \bibinfo {author}
  {\bibfnamefont {M.~A.}\ \bibnamefont {Eriksson}},\ }\href
  {https://doi.org/10.1038/s41467-022-35510-z} {\bibfield  {journal} {\bibinfo
  {journal} {Nat. Commun.}\ }\textbf {\bibinfo {volume} {13}},\ \bibinfo
  {pages} {7777} (\bibinfo {year} {2022})}\BibitemShut {NoStop}%
\bibitem [{\citenamefont {{Paquelet Wuetz}}\ \emph {et~al.}(2022)\citenamefont
  {{Paquelet Wuetz}}, \citenamefont {Losert}, \citenamefont {Koelling},
  \citenamefont {Stehouwer}, \citenamefont {Zwerver}, \citenamefont {Philips},
  \citenamefont {M{\c{a}}dzik}, \citenamefont {Xue}, \citenamefont {Zheng},
  \citenamefont {Lodari}, \citenamefont {Amitonov}, \citenamefont
  {Samkharadze}, \citenamefont {Sammak}, \citenamefont {Vandersypen},
  \citenamefont {Rahman}, \citenamefont {Coppersmith}, \citenamefont
  {Moutanabbir}, \citenamefont {Friesen},\ and\ \citenamefont
  {Scappucci}}]{PaqueletWuetz2022}%
  \BibitemOpen
  \bibfield  {author} {\bibinfo {author} {\bibfnamefont {B.}~\bibnamefont
  {{Paquelet Wuetz}}}, \bibinfo {author} {\bibfnamefont {M.~P.}\ \bibnamefont
  {Losert}}, \bibinfo {author} {\bibfnamefont {S.}~\bibnamefont {Koelling}},
  \bibinfo {author} {\bibfnamefont {L.~E.}\ \bibnamefont {Stehouwer}}, \bibinfo
  {author} {\bibfnamefont {A.~M.~J.}\ \bibnamefont {Zwerver}}, \bibinfo
  {author} {\bibfnamefont {S.~G.}\ \bibnamefont {Philips}}, \bibinfo {author}
  {\bibfnamefont {M.~T.}\ \bibnamefont {M{\c{a}}dzik}}, \bibinfo {author}
  {\bibfnamefont {X.}~\bibnamefont {Xue}}, \bibinfo {author} {\bibfnamefont
  {G.}~\bibnamefont {Zheng}}, \bibinfo {author} {\bibfnamefont
  {M.}~\bibnamefont {Lodari}}, \bibinfo {author} {\bibfnamefont {S.~V.}\
  \bibnamefont {Amitonov}}, \bibinfo {author} {\bibfnamefont {N.}~\bibnamefont
  {Samkharadze}}, \bibinfo {author} {\bibfnamefont {A.}~\bibnamefont {Sammak}},
  \bibinfo {author} {\bibfnamefont {L.~M.}\ \bibnamefont {Vandersypen}},
  \bibinfo {author} {\bibfnamefont {R.}~\bibnamefont {Rahman}}, \bibinfo
  {author} {\bibfnamefont {S.~N.}\ \bibnamefont {Coppersmith}}, \bibinfo
  {author} {\bibfnamefont {O.}~\bibnamefont {Moutanabbir}}, \bibinfo {author}
  {\bibfnamefont {M.}~\bibnamefont {Friesen}},\ and\ \bibinfo {author}
  {\bibfnamefont {G.}~\bibnamefont {Scappucci}},\ }\href
  {https://doi.org/10.1038/s41467-022-35458-0} {\bibfield  {journal} {\bibinfo
  {journal} {Nat. Commun.}\ }\textbf {\bibinfo {volume} {13}},\ \bibinfo
  {pages} {7730} (\bibinfo {year} {2022})}\BibitemShut {NoStop}%
\bibitem [{\citenamefont {Vandersypen}\ \emph {et~al.}(2017)\citenamefont
  {Vandersypen}, \citenamefont {Bluhm}, \citenamefont {Clarke}, \citenamefont
  {Dzurak}, \citenamefont {Ishihara}, \citenamefont {Morello}, \citenamefont
  {Reilly}, \citenamefont {Schreiber},\ and\ \citenamefont
  {Veldhorst}}]{Vandersypen2017}%
  \BibitemOpen
  \bibfield  {author} {\bibinfo {author} {\bibfnamefont {L.~M.}\ \bibnamefont
  {Vandersypen}}, \bibinfo {author} {\bibfnamefont {H.}~\bibnamefont {Bluhm}},
  \bibinfo {author} {\bibfnamefont {J.~S.}\ \bibnamefont {Clarke}}, \bibinfo
  {author} {\bibfnamefont {A.~S.}\ \bibnamefont {Dzurak}}, \bibinfo {author}
  {\bibfnamefont {R.}~\bibnamefont {Ishihara}}, \bibinfo {author}
  {\bibfnamefont {A.}~\bibnamefont {Morello}}, \bibinfo {author} {\bibfnamefont
  {D.~J.}\ \bibnamefont {Reilly}}, \bibinfo {author} {\bibfnamefont {L.~R.}\
  \bibnamefont {Schreiber}},\ and\ \bibinfo {author} {\bibfnamefont
  {M.}~\bibnamefont {Veldhorst}},\ }\href
  {https://doi.org/10.1038/S41534-017-0038-Y} {\bibfield  {journal} {\bibinfo
  {journal} {npj Quantum Inf.}\ }\textbf {\bibinfo {volume} {3}},\ \bibinfo
  {pages} {34} (\bibinfo {year} {2017})}\BibitemShut {NoStop}%
\bibitem [{\citenamefont {Chen}\ \emph {et~al.}(2021)\citenamefont {Chen},
  \citenamefont {Raach}, \citenamefont {Pan}, \citenamefont {Kiselev},
  \citenamefont {Acuna}, \citenamefont {Blumoff}, \citenamefont {Brecht},
  \citenamefont {Choi}, \citenamefont {Ha}, \citenamefont {Hulbert},
  \citenamefont {Jura}, \citenamefont {Keating}, \citenamefont {Noah},
  \citenamefont {Sun}, \citenamefont {Thomas}, \citenamefont {Borselli},
  \citenamefont {Jackson}, \citenamefont {Rakher},\ and\ \citenamefont
  {Ross}}]{Chen2021}%
  \BibitemOpen
  \bibfield  {author} {\bibinfo {author} {\bibfnamefont {E.~H.}\ \bibnamefont
  {Chen}}, \bibinfo {author} {\bibfnamefont {K.}~\bibnamefont {Raach}},
  \bibinfo {author} {\bibfnamefont {A.}~\bibnamefont {Pan}}, \bibinfo {author}
  {\bibfnamefont {A.~A.}\ \bibnamefont {Kiselev}}, \bibinfo {author}
  {\bibfnamefont {E.}~\bibnamefont {Acuna}}, \bibinfo {author} {\bibfnamefont
  {J.~Z.}\ \bibnamefont {Blumoff}}, \bibinfo {author} {\bibfnamefont
  {T.}~\bibnamefont {Brecht}}, \bibinfo {author} {\bibfnamefont {M.~D.}\
  \bibnamefont {Choi}}, \bibinfo {author} {\bibfnamefont {W.}~\bibnamefont
  {Ha}}, \bibinfo {author} {\bibfnamefont {D.~R.}\ \bibnamefont {Hulbert}},
  \bibinfo {author} {\bibfnamefont {M.~P.}\ \bibnamefont {Jura}}, \bibinfo
  {author} {\bibfnamefont {T.~E.}\ \bibnamefont {Keating}}, \bibinfo {author}
  {\bibfnamefont {R.}~\bibnamefont {Noah}}, \bibinfo {author} {\bibfnamefont
  {B.}~\bibnamefont {Sun}}, \bibinfo {author} {\bibfnamefont {B.~J.}\
  \bibnamefont {Thomas}}, \bibinfo {author} {\bibfnamefont {M.~G.}\
  \bibnamefont {Borselli}}, \bibinfo {author} {\bibfnamefont {C.~A.~C.}\
  \bibnamefont {Jackson}}, \bibinfo {author} {\bibfnamefont {M.~T.}\
  \bibnamefont {Rakher}},\ and\ \bibinfo {author} {\bibfnamefont {R.~S.}\
  \bibnamefont {Ross}},\ }\href
  {https://doi.org/10.1103/PhysRevApplied.15.044033} {\bibfield  {journal}
  {\bibinfo  {journal} {Phys. Rev. App.}\ }\textbf {\bibinfo {volume} {15}},\
  \bibinfo {pages} {044033} (\bibinfo {year} {2021})}\BibitemShut {NoStop}%
\bibitem [{\citenamefont {Langrock}\ \emph {et~al.}(2023)\citenamefont
  {Langrock}, \citenamefont {Krzywda}, \citenamefont {Focke}, \citenamefont
  {Seidler}, \citenamefont {Schreiber},\ and\ \citenamefont
  {Cywi{\'{n}}ski}}]{Langrock2022}%
  \BibitemOpen
  \bibfield  {author} {\bibinfo {author} {\bibfnamefont {V.}~\bibnamefont
  {Langrock}}, \bibinfo {author} {\bibfnamefont {J.~A.}\ \bibnamefont
  {Krzywda}}, \bibinfo {author} {\bibfnamefont {N.}~\bibnamefont {Focke}},
  \bibinfo {author} {\bibfnamefont {I.}~\bibnamefont {Seidler}}, \bibinfo
  {author} {\bibfnamefont {L.~R.}\ \bibnamefont {Schreiber}},\ and\ \bibinfo
  {author} {\bibfnamefont {{\L}.}~\bibnamefont {Cywi{\'{n}}ski}},\ }\href
  {https://doi.org/10.1103/PRXQuantum.4.020305} {\bibfield  {journal} {\bibinfo
   {journal} {PRX Quantum}\ }\textbf {\bibinfo {volume} {4}},\ \bibinfo {pages}
  {020305} (\bibinfo {year} {2023})}\BibitemShut {NoStop}%
\bibitem [{\citenamefont {Beckers}\ \emph {et~al.}(2020)\citenamefont
  {Beckers}, \citenamefont {Jazaeri}, \citenamefont {Grill}, \citenamefont
  {Narasimhamoorthy}, \citenamefont {Parvais},\ and\ \citenamefont
  {Enz}}]{Beckers2020}%
  \BibitemOpen
  \bibfield  {author} {\bibinfo {author} {\bibfnamefont {A.}~\bibnamefont
  {Beckers}}, \bibinfo {author} {\bibfnamefont {F.}~\bibnamefont {Jazaeri}},
  \bibinfo {author} {\bibfnamefont {A.}~\bibnamefont {Grill}}, \bibinfo
  {author} {\bibfnamefont {S.}~\bibnamefont {Narasimhamoorthy}}, \bibinfo
  {author} {\bibfnamefont {B.}~\bibnamefont {Parvais}},\ and\ \bibinfo {author}
  {\bibfnamefont {C.}~\bibnamefont {Enz}},\ }\href
  {https://doi.org/10.1109/JEDS.2020.2989629} {\bibfield  {journal} {\bibinfo
  {journal} {IEEE J-EDS}\ }\textbf {\bibinfo {volume} {8}},\ \bibinfo {pages}
  {780} (\bibinfo {year} {2020})}\BibitemShut {NoStop}%
\bibitem [{\citenamefont {Rimini}(1995)}]{Rimini1995}%
  \BibitemOpen
  \bibfield  {author} {\bibinfo {author} {\bibfnamefont {E.}~\bibnamefont
  {Rimini}},\ }\href@noop {} {\emph {\bibinfo {title} {{Ion Implantation:
  basics to device fabrication}}}}\ (\bibinfo  {publisher} {Springer
  Science+Business Media, LLC},\ \bibinfo {year} {1995})\BibitemShut {NoStop}%
\bibitem [{\citenamefont {Lawrie}\ \emph {et~al.}(2020)\citenamefont {Lawrie},
  \citenamefont {Eenink}, \citenamefont {Hendrickx}, \citenamefont {Boter},
  \citenamefont {Petit}, \citenamefont {Amitonov}, \citenamefont {Lodari},
  \citenamefont {{Paquelet Wuetz}}, \citenamefont {Volk}, \citenamefont
  {Philips}, \citenamefont {Droulers}, \citenamefont {Kalhor}, \citenamefont
  {{Van Riggelen}}, \citenamefont {Brousse}, \citenamefont {Sammak},
  \citenamefont {Vandersypen}, \citenamefont {Scappucci},\ and\ \citenamefont
  {Veldhorst}}]{Lawrie2020}%
  \BibitemOpen
  \bibfield  {author} {\bibinfo {author} {\bibfnamefont {W.~I.}\ \bibnamefont
  {Lawrie}}, \bibinfo {author} {\bibfnamefont {H.~G.}\ \bibnamefont {Eenink}},
  \bibinfo {author} {\bibfnamefont {N.~W.}\ \bibnamefont {Hendrickx}}, \bibinfo
  {author} {\bibfnamefont {J.~M.}\ \bibnamefont {Boter}}, \bibinfo {author}
  {\bibfnamefont {L.}~\bibnamefont {Petit}}, \bibinfo {author} {\bibfnamefont
  {S.~V.}\ \bibnamefont {Amitonov}}, \bibinfo {author} {\bibfnamefont
  {M.}~\bibnamefont {Lodari}}, \bibinfo {author} {\bibfnamefont
  {B.}~\bibnamefont {{Paquelet Wuetz}}}, \bibinfo {author} {\bibfnamefont
  {C.}~\bibnamefont {Volk}}, \bibinfo {author} {\bibfnamefont {S.~G.}\
  \bibnamefont {Philips}}, \bibinfo {author} {\bibfnamefont {G.}~\bibnamefont
  {Droulers}}, \bibinfo {author} {\bibfnamefont {N.}~\bibnamefont {Kalhor}},
  \bibinfo {author} {\bibfnamefont {F.}~\bibnamefont {{Van Riggelen}}},
  \bibinfo {author} {\bibfnamefont {D.}~\bibnamefont {Brousse}}, \bibinfo
  {author} {\bibfnamefont {A.}~\bibnamefont {Sammak}}, \bibinfo {author}
  {\bibfnamefont {L.~M.}\ \bibnamefont {Vandersypen}}, \bibinfo {author}
  {\bibfnamefont {G.}~\bibnamefont {Scappucci}},\ and\ \bibinfo {author}
  {\bibfnamefont {M.}~\bibnamefont {Veldhorst}},\ }\href
  {https://doi.org/10.1063/5.0002013} {\bibfield  {journal} {\bibinfo
  {journal} {Appl. Phys. Lett.}\ }\textbf {\bibinfo {volume} {116}},\ \bibinfo
  {pages} {080501} (\bibinfo {year} {2020})}\BibitemShut {NoStop}%
\bibitem [{\citenamefont {Seidler}\ \emph {et~al.}(2022)\citenamefont
  {Seidler}, \citenamefont {Struck}, \citenamefont {Xue}, \citenamefont
  {Focke}, \citenamefont {Trellenkamp}, \citenamefont {Bluhm},\ and\
  \citenamefont {Schreiber}}]{Seidler2022}%
  \BibitemOpen
  \bibfield  {author} {\bibinfo {author} {\bibfnamefont {I.}~\bibnamefont
  {Seidler}}, \bibinfo {author} {\bibfnamefont {T.}~\bibnamefont {Struck}},
  \bibinfo {author} {\bibfnamefont {R.}~\bibnamefont {Xue}}, \bibinfo {author}
  {\bibfnamefont {N.}~\bibnamefont {Focke}}, \bibinfo {author} {\bibfnamefont
  {S.}~\bibnamefont {Trellenkamp}}, \bibinfo {author} {\bibfnamefont
  {H.}~\bibnamefont {Bluhm}},\ and\ \bibinfo {author} {\bibfnamefont {L.~R.}\
  \bibnamefont {Schreiber}},\ }\href
  {https://doi.org/10.1038/s41534-022-00615-2} {\bibfield  {journal} {\bibinfo
  {journal} {npj Quantum Inf.}\ }\textbf {\bibinfo {volume} {8}},\ \bibinfo
  {pages} {100} (\bibinfo {year} {2022})}\BibitemShut {NoStop}%
\bibitem [{\citenamefont {Taiwo}\ \emph {et~al.}(2022)\citenamefont {Taiwo},
  \citenamefont {Shin},\ and\ \citenamefont {Son}}]{Taiwo2022}%
  \BibitemOpen
  \bibfield  {author} {\bibinfo {author} {\bibfnamefont {R.~A.}\ \bibnamefont
  {Taiwo}}, \bibinfo {author} {\bibfnamefont {J.-h.}\ \bibnamefont {Shin}},\
  and\ \bibinfo {author} {\bibfnamefont {Y.-I.}\ \bibnamefont {Son}},\ }\href
  {https://doi.org/https://doi.org/10.3390/ ma15227886} {\bibfield  {journal}
  {\bibinfo  {journal} {Materials}\ }\textbf {\bibinfo {volume} {15}},\
  \bibinfo {pages} {7886} (\bibinfo {year} {2022})}\BibitemShut {NoStop}%
\bibitem [{\citenamefont {Tabata}\ \emph
  {et~al.}(2022{\natexlab{a}})\citenamefont {Tabata}, \citenamefont {Roze},
  \citenamefont {Alba}, \citenamefont {Halty}, \citenamefont {Raynal},
  \citenamefont {Karmous}, \citenamefont {Kerdiles},\ and\ \citenamefont
  {Mazzamuto}}]{Tabata2022}%
  \BibitemOpen
  \bibfield  {author} {\bibinfo {author} {\bibfnamefont {T.}~\bibnamefont
  {Tabata}}, \bibinfo {author} {\bibfnamefont {F.}~\bibnamefont {Roze}},
  \bibinfo {author} {\bibfnamefont {P.~A.}\ \bibnamefont {Alba}}, \bibinfo
  {author} {\bibfnamefont {S.}~\bibnamefont {Halty}}, \bibinfo {author}
  {\bibfnamefont {P.~E.}\ \bibnamefont {Raynal}}, \bibinfo {author}
  {\bibfnamefont {I.}~\bibnamefont {Karmous}}, \bibinfo {author} {\bibfnamefont
  {S.}~\bibnamefont {Kerdiles}},\ and\ \bibinfo {author} {\bibfnamefont
  {F.}~\bibnamefont {Mazzamuto}},\ }\href
  {https://doi.org/10.1109/JEDS.2021.3131911} {\bibfield  {journal} {\bibinfo
  {journal} {IEEE J-EDS}\ }\textbf {\bibinfo {volume} {10}},\ \bibinfo {pages}
  {712} (\bibinfo {year} {2022}{\natexlab{a}})}\BibitemShut {NoStop}%
\bibitem [{\citenamefont {Tabata}\ \emph
  {et~al.}(2022{\natexlab{b}})\citenamefont {Tabata}, \citenamefont
  {Roz{\'{e}}}, \citenamefont {Thuries}, \citenamefont {Halty}, \citenamefont
  {Raynal}, \citenamefont {Huet}, \citenamefont {Mazzamuto}, \citenamefont
  {Joshi},\ and\ \citenamefont {Basol}}]{Tabata2022a}%
  \BibitemOpen
  \bibfield  {author} {\bibinfo {author} {\bibfnamefont {T.}~\bibnamefont
  {Tabata}}, \bibinfo {author} {\bibfnamefont {F.}~\bibnamefont {Roz{\'{e}}}},
  \bibinfo {author} {\bibfnamefont {L.}~\bibnamefont {Thuries}}, \bibinfo
  {author} {\bibfnamefont {S.}~\bibnamefont {Halty}}, \bibinfo {author}
  {\bibfnamefont {P.-e.}\ \bibnamefont {Raynal}}, \bibinfo {author}
  {\bibfnamefont {K.}~\bibnamefont {Huet}}, \bibinfo {author} {\bibfnamefont
  {F.}~\bibnamefont {Mazzamuto}}, \bibinfo {author} {\bibfnamefont
  {A.}~\bibnamefont {Joshi}},\ and\ \bibinfo {author} {\bibfnamefont {B.~M.}\
  \bibnamefont {Basol}},\ }\href {https://doi.org/10.35848/1882-0786/ac6e2a}
  {\bibfield  {journal} {\bibinfo  {journal} {Appl. Phys. Express}\ }\textbf
  {\bibinfo {volume} {15}},\ \bibinfo {pages} {061002} (\bibinfo {year}
  {2022}{\natexlab{b}})}\BibitemShut {NoStop}%
\bibitem [{\citenamefont {Mishima}\ \emph {et~al.}(2005)\citenamefont
  {Mishima}, \citenamefont {Ochimizu},\ and\ \citenamefont
  {Mimura}}]{Mishima2005}%
  \BibitemOpen
  \bibfield  {author} {\bibinfo {author} {\bibfnamefont {Y.}~\bibnamefont
  {Mishima}}, \bibinfo {author} {\bibfnamefont {H.}~\bibnamefont {Ochimizu}},\
  and\ \bibinfo {author} {\bibfnamefont {A.}~\bibnamefont {Mimura}},\ }\href
  {https://doi.org/10.1143/JJAP.44.2336} {\bibfield  {journal} {\bibinfo
  {journal} {Jpn. J. Appl. Phys.}\ }\textbf {\bibinfo {volume} {44}},\ \bibinfo
  {pages} {2336} (\bibinfo {year} {2005})}\BibitemShut {NoStop}%
\bibitem [{\citenamefont {Kim}\ \emph {et~al.}(2022)\citenamefont {Kim},
  \citenamefont {Junger}, \citenamefont {Morvan}, \citenamefont {Barnard},
  \citenamefont {Livingston}, \citenamefont {Altoe}, \citenamefont {Kim},
  \citenamefont {Song}, \citenamefont {Chen}, \citenamefont {Kreikebaum},
  \citenamefont {Ogletree}, \citenamefont {Santiago},\ and\ \citenamefont
  {Siddiqi}}]{Kim2022}%
  \BibitemOpen
  \bibfield  {author} {\bibinfo {author} {\bibfnamefont {H.}~\bibnamefont
  {Kim}}, \bibinfo {author} {\bibfnamefont {C.}~\bibnamefont {Junger}},
  \bibinfo {author} {\bibfnamefont {A.}~\bibnamefont {Morvan}}, \bibinfo
  {author} {\bibfnamefont {E.~S.}\ \bibnamefont {Barnard}}, \bibinfo {author}
  {\bibfnamefont {W.~P.}\ \bibnamefont {Livingston}}, \bibinfo {author}
  {\bibfnamefont {M.~V.~P.}\ \bibnamefont {Altoe}}, \bibinfo {author}
  {\bibfnamefont {Y.}~\bibnamefont {Kim}}, \bibinfo {author} {\bibfnamefont
  {C.}~\bibnamefont {Song}}, \bibinfo {author} {\bibfnamefont {L.}~\bibnamefont
  {Chen}}, \bibinfo {author} {\bibfnamefont {J.~M.}\ \bibnamefont
  {Kreikebaum}}, \bibinfo {author} {\bibfnamefont {D.~F.}\ \bibnamefont
  {Ogletree}}, \bibinfo {author} {\bibfnamefont {D.~I.}\ \bibnamefont
  {Santiago}},\ and\ \bibinfo {author} {\bibfnamefont {I.}~\bibnamefont
  {Siddiqi}},\ }\href {https://doi.org/10.1063/5.0102092} {\bibfield  {journal}
  {\bibinfo  {journal} {Appl. Phys. Lett.}\ }\textbf {\bibinfo {volume}
  {121}},\ \bibinfo {pages} {142601} (\bibinfo {year} {2022})}\BibitemShut
  {NoStop}%
\bibitem [{\citenamefont {Ziegler}\ and\ \citenamefont
  {Biersack}(1985)}]{Bromley1985}%
  \BibitemOpen
  \bibfield  {author} {\bibinfo {author} {\bibfnamefont {J.~F.}\ \bibnamefont
  {Ziegler}}\ and\ \bibinfo {author} {\bibfnamefont {J.~P.}\ \bibnamefont
  {Biersack}},\ }in\ \href@noop {} {\emph {\bibinfo {booktitle} {Treatise on
  Heavy-Ion Science}}}\ (\bibinfo  {publisher} {Springer Science+Business Media
  New York},\ \bibinfo {year} {1985})\ Chap.~\bibinfo {chapter} {3}, pp.\
  \bibinfo {pages} {95--129}\BibitemShut {NoStop}%
\bibitem [{\citenamefont {Wesch}\ and\ \citenamefont
  {Goetz}(1980)}]{Wesch1980}%
  \BibitemOpen
  \bibfield  {author} {\bibinfo {author} {\bibfnamefont {W.}~\bibnamefont
  {Wesch}}\ and\ \bibinfo {author} {\bibfnamefont {G.}~\bibnamefont {Goetz}},\
  }\href {https://doi.org/10.1080/00337578008243082} {\bibfield  {journal}
  {\bibinfo  {journal} {Radiation effects}\ }\textbf {\bibinfo {volume} {49}},\
  \bibinfo {pages} {137} (\bibinfo {year} {1980})}\BibitemShut {NoStop}%
\bibitem [{\citenamefont {Danz}\ and\ \citenamefont
  {Gretscher}(2004)}]{Danz2004}%
  \BibitemOpen
  \bibfield  {author} {\bibinfo {author} {\bibfnamefont {R.}~\bibnamefont
  {Danz}}\ and\ \bibinfo {author} {\bibfnamefont {P.}~\bibnamefont
  {Gretscher}},\ }\href {https://doi.org/10.1016/j.tsf.2004.05.124} {\bibfield
  {journal} {\bibinfo  {journal} {Thin Solid Films}\ }\textbf {\bibinfo
  {volume} {462-463}},\ \bibinfo {pages} {257} (\bibinfo {year}
  {2004})}\BibitemShut {NoStop}%
\bibitem [{\citenamefont {Chen}\ \emph {et~al.}(2002)\citenamefont {Chen},
  \citenamefont {Li}, \citenamefont {Peng}, \citenamefont {Liu}, \citenamefont
  {Liu}, \citenamefont {Huang}, \citenamefont {Zhou},\ and\ \citenamefont
  {Xue}}]{Chen2002}%
  \BibitemOpen
  \bibfield  {author} {\bibinfo {author} {\bibfnamefont {H.}~\bibnamefont
  {Chen}}, \bibinfo {author} {\bibfnamefont {Y.~K.}\ \bibnamefont {Li}},
  \bibinfo {author} {\bibfnamefont {C.~S.}\ \bibnamefont {Peng}}, \bibinfo
  {author} {\bibfnamefont {H.~F.}\ \bibnamefont {Liu}}, \bibinfo {author}
  {\bibfnamefont {Y.~L.}\ \bibnamefont {Liu}}, \bibinfo {author} {\bibfnamefont
  {Q.}~\bibnamefont {Huang}}, \bibinfo {author} {\bibfnamefont {J.~M.}\
  \bibnamefont {Zhou}},\ and\ \bibinfo {author} {\bibfnamefont {Q.~K.}\
  \bibnamefont {Xue}},\ }\href {https://doi.org/10.1103/PhysRevB.65.233303}
  {\bibfield  {journal} {\bibinfo  {journal} {Phys. Rev. B}\ }\textbf {\bibinfo
  {volume} {65}},\ \bibinfo {pages} {233303} (\bibinfo {year}
  {2002})}\BibitemShut {NoStop}%
\bibitem [{\citenamefont {Auston}\ \emph {et~al.}(1979)\citenamefont {Auston},
  \citenamefont {Golovchenko}, \citenamefont {Simons}, \citenamefont {Slusher},
  \citenamefont {Smith}, \citenamefont {Surko},\ and\ \citenamefont
  {Venkatesan}}]{Marine1983}%
  \BibitemOpen
  \bibfield  {author} {\bibinfo {author} {\bibfnamefont {D.~H.}\ \bibnamefont
  {Auston}}, \bibinfo {author} {\bibfnamefont {J.~A.}\ \bibnamefont
  {Golovchenko}}, \bibinfo {author} {\bibfnamefont {A.~L.}\ \bibnamefont
  {Simons}}, \bibinfo {author} {\bibfnamefont {R.~E.}\ \bibnamefont {Slusher}},
  \bibinfo {author} {\bibfnamefont {P.~R.}\ \bibnamefont {Smith}}, \bibinfo
  {author} {\bibfnamefont {C.~M.}\ \bibnamefont {Surko}},\ and\ \bibinfo
  {author} {\bibfnamefont {T.~N.~C.}\ \bibnamefont {Venkatesan}},\ }\href
  {https://doi.org/10.1051/jphyscol:1983520} {\bibfield  {journal} {\bibinfo
  {journal} {AIP Conf.}\ }\textbf {\bibinfo {volume} {50}},\ \bibinfo {pages}
  {11} (\bibinfo {year} {1979})}\BibitemShut {NoStop}%
\bibitem [{\citenamefont {Boyd}\ and\ \citenamefont {Wilson}(1983)}]{Boyd1983}%
  \BibitemOpen
  \bibfield  {author} {\bibinfo {author} {\bibfnamefont {I.~W.}\ \bibnamefont
  {Boyd}}\ and\ \bibinfo {author} {\bibfnamefont {J.~I.}\ \bibnamefont
  {Wilson}},\ }\href {https://doi.org/10.1038/303481a0} {\bibfield  {journal}
  {\bibinfo  {journal} {Nature}\ }\textbf {\bibinfo {volume} {303}},\ \bibinfo
  {pages} {481} (\bibinfo {year} {1983})}\BibitemShut {NoStop}%
\bibitem [{\citenamefont {Olesinski}\ and\ \citenamefont
  {Abbaschian}(1984)}]{Olesinski1984}%
  \BibitemOpen
  \bibfield  {author} {\bibinfo {author} {\bibfnamefont {R.~W.}\ \bibnamefont
  {Olesinski}}\ and\ \bibinfo {author} {\bibfnamefont {G.~J.}\ \bibnamefont
  {Abbaschian}},\ }\href {https://doi.org/https://doi.org/10.1007/BF02868957}
  {\bibfield  {journal} {\bibinfo  {journal} {Bull. alloy phase diagr.}\
  }\textbf {\bibinfo {volume} {5}},\ \bibinfo {pages} {180} (\bibinfo {year}
  {1984})}\BibitemShut {NoStop}%
\bibitem [{\citenamefont {Luong}\ \emph {et~al.}(2013)\citenamefont {Luong},
  \citenamefont {Wirths}, \citenamefont {Stefanov}, \citenamefont
  {Holl{\"{a}}nder}, \citenamefont {Schubert}, \citenamefont {Conde},
  \citenamefont {Stoica}, \citenamefont {Breuer}, \citenamefont {Chiussi},
  \citenamefont {Goryll}, \citenamefont {Buca},\ and\ \citenamefont
  {Mantl}}]{Luong2013}%
  \BibitemOpen
  \bibfield  {author} {\bibinfo {author} {\bibfnamefont {G.~V.}\ \bibnamefont
  {Luong}}, \bibinfo {author} {\bibfnamefont {S.}~\bibnamefont {Wirths}},
  \bibinfo {author} {\bibfnamefont {S.}~\bibnamefont {Stefanov}}, \bibinfo
  {author} {\bibfnamefont {B.}~\bibnamefont {Holl{\"{a}}nder}}, \bibinfo
  {author} {\bibfnamefont {J.}~\bibnamefont {Schubert}}, \bibinfo {author}
  {\bibfnamefont {J.~C.}\ \bibnamefont {Conde}}, \bibinfo {author}
  {\bibfnamefont {T.}~\bibnamefont {Stoica}}, \bibinfo {author} {\bibfnamefont
  {U.}~\bibnamefont {Breuer}}, \bibinfo {author} {\bibfnamefont
  {S.}~\bibnamefont {Chiussi}}, \bibinfo {author} {\bibfnamefont
  {M.}~\bibnamefont {Goryll}}, \bibinfo {author} {\bibfnamefont
  {D.}~\bibnamefont {Buca}},\ and\ \bibinfo {author} {\bibfnamefont
  {S.}~\bibnamefont {Mantl}},\ }\href {https://doi.org/10.1063/1.4807001}
  {\bibfield  {journal} {\bibinfo  {journal} {J. Appl. Phys.}\ }\textbf
  {\bibinfo {volume} {113}},\ \bibinfo {pages} {204902} (\bibinfo {year}
  {2013})}\BibitemShut {NoStop}%
\bibitem [{\citenamefont {Lee}\ \emph {et~al.}(1997)\citenamefont {Lee},
  \citenamefont {Cahill},\ and\ \citenamefont {Venkatasubramianian}}]{Lee1997}%
  \BibitemOpen
  \bibfield  {author} {\bibinfo {author} {\bibfnamefont {S.-M.}\ \bibnamefont
  {Lee}}, \bibinfo {author} {\bibfnamefont {D.}~\bibnamefont {Cahill}},\ and\
  \bibinfo {author} {\bibfnamefont {R.}~\bibnamefont {Venkatasubramianian}},\
  }\href {https://doi.org/https://doi.org/10.1063/1.118755} {\bibfield
  {journal} {\bibinfo  {journal} {Appl. Phys. Lett.}\ }\textbf {\bibinfo
  {volume} {77}},\ \bibinfo {pages} {2957} (\bibinfo {year}
  {1997})}\BibitemShut {NoStop}%
\bibitem [{\citenamefont {Paul}(2004)}]{Paul2004}%
  \BibitemOpen
  \bibfield  {author} {\bibinfo {author} {\bibfnamefont {D.~J.}\ \bibnamefont
  {Paul}},\ }\href {https://doi.org/10.1088/0268-1242/19/10/R02} {\bibfield
  {journal} {\bibinfo  {journal} {Semicond. Sci. Technol.}\ }\textbf {\bibinfo
  {volume} {19}},\ \bibinfo {pages} {75} (\bibinfo {year} {2004})}\BibitemShut
  {NoStop}%
\bibitem [{\citenamefont {Huang}\ \emph {et~al.}(2014)\citenamefont {Huang},
  \citenamefont {Li}, \citenamefont {Chou},\ and\ \citenamefont
  {Sturm}}]{Huang2014}%
  \BibitemOpen
  \bibfield  {author} {\bibinfo {author} {\bibfnamefont {C.~T.}\ \bibnamefont
  {Huang}}, \bibinfo {author} {\bibfnamefont {J.~Y.}\ \bibnamefont {Li}},
  \bibinfo {author} {\bibfnamefont {K.~S.}\ \bibnamefont {Chou}},\ and\
  \bibinfo {author} {\bibfnamefont {J.~C.}\ \bibnamefont {Sturm}},\ }\href
  {https://doi.org/10.1063/1.4884650} {\bibfield  {journal} {\bibinfo
  {journal} {Appl. Phys. Lett.}\ }\textbf {\bibinfo {volume} {104}},\ \bibinfo
  {pages} {243510} (\bibinfo {year} {2014})}\BibitemShut {NoStop}%
\bibitem [{\citenamefont {Laroche}\ \emph {et~al.}(2015)\citenamefont
  {Laroche}, \citenamefont {Huang}, \citenamefont {Nielsen}, \citenamefont
  {Chuang}, \citenamefont {Li}, \citenamefont {Liu},\ and\ \citenamefont
  {Lu}}]{Laroche2015}%
  \BibitemOpen
  \bibfield  {author} {\bibinfo {author} {\bibfnamefont {D.}~\bibnamefont
  {Laroche}}, \bibinfo {author} {\bibfnamefont {S.~H.}\ \bibnamefont {Huang}},
  \bibinfo {author} {\bibfnamefont {E.}~\bibnamefont {Nielsen}}, \bibinfo
  {author} {\bibfnamefont {Y.}~\bibnamefont {Chuang}}, \bibinfo {author}
  {\bibfnamefont {J.~Y.}\ \bibnamefont {Li}}, \bibinfo {author} {\bibfnamefont
  {C.~W.}\ \bibnamefont {Liu}},\ and\ \bibinfo {author} {\bibfnamefont {T.~M.}\
  \bibnamefont {Lu}},\ }\href {https://doi.org/10.1063/1.4933026} {\bibfield
  {journal} {\bibinfo  {journal} {AIP Adv.}\ }\textbf {\bibinfo {volume} {5}},\
  \bibinfo {pages} {107106} (\bibinfo {year} {2015})}\BibitemShut {NoStop}%
\bibitem [{\citenamefont {Mi}\ \emph {et~al.}(2015)\citenamefont {Mi},
  \citenamefont {Hazard}, \citenamefont {Payette}, \citenamefont {Wang},
  \citenamefont {Zajac}, \citenamefont {Cady},\ and\ \citenamefont
  {Petta}}]{Mi2015}%
  \BibitemOpen
  \bibfield  {author} {\bibinfo {author} {\bibfnamefont {X.}~\bibnamefont
  {Mi}}, \bibinfo {author} {\bibfnamefont {T.~M.}\ \bibnamefont {Hazard}},
  \bibinfo {author} {\bibfnamefont {C.}~\bibnamefont {Payette}}, \bibinfo
  {author} {\bibfnamefont {K.}~\bibnamefont {Wang}}, \bibinfo {author}
  {\bibfnamefont {D.~M.}\ \bibnamefont {Zajac}}, \bibinfo {author}
  {\bibfnamefont {J.~V.}\ \bibnamefont {Cady}},\ and\ \bibinfo {author}
  {\bibfnamefont {J.~R.}\ \bibnamefont {Petta}},\ }\href
  {https://doi.org/10.1103/PhysRevB.92.035304} {\bibfield  {journal} {\bibinfo
  {journal} {Phys. Rev. Bhy}\ }\textbf {\bibinfo {volume} {92}},\ \bibinfo
  {pages} {035304} (\bibinfo {year} {2015})}\BibitemShut {NoStop}%
\bibitem [{\citenamefont {Monroe}(1993)}]{Monroe1993}%
  \BibitemOpen
  \bibfield  {author} {\bibinfo {author} {\bibfnamefont {D.}~\bibnamefont
  {Monroe}},\ }\href {https://doi.org/10.1116/1.586471} {\bibfield  {journal}
  {\bibinfo  {journal} {J. Vac. Sci. Technol. B.}\ }\textbf {\bibinfo {volume}
  {11}},\ \bibinfo {pages} {1731} (\bibinfo {year} {1993})}\BibitemShut
  {NoStop}%
\bibitem [{\citenamefont {{Das Sarma}}\ and\ \citenamefont
  {Hwang}(2013)}]{DasSarma2013}%
  \BibitemOpen
  \bibfield  {author} {\bibinfo {author} {\bibfnamefont {S.}~\bibnamefont {{Das
  Sarma}}}\ and\ \bibinfo {author} {\bibfnamefont {E.~H.}\ \bibnamefont
  {Hwang}},\ }\href {https://doi.org/10.1103/PhysRevB.88.035439} {\bibfield
  {journal} {\bibinfo  {journal} {Phys. Rev. B}\ }\textbf {\bibinfo {volume}
  {88}},\ \bibinfo {pages} {035439} (\bibinfo {year} {2013})}\BibitemShut
  {NoStop}%
\bibitem [{\citenamefont {Dolgopolov}\ \emph {et~al.}(2003)\citenamefont
  {Dolgopolov}, \citenamefont {Deviatov}, \citenamefont {Shashkin},
  \citenamefont {Wieser}, \citenamefont {Kunze}, \citenamefont {Abstreiter},\
  and\ \citenamefont {Brunner}}]{Dolgopolov2003}%
  \BibitemOpen
  \bibfield  {author} {\bibinfo {author} {\bibfnamefont {V.~T.}\ \bibnamefont
  {Dolgopolov}}, \bibinfo {author} {\bibfnamefont {E.~V.}\ \bibnamefont
  {Deviatov}}, \bibinfo {author} {\bibfnamefont {A.~A.}\ \bibnamefont
  {Shashkin}}, \bibinfo {author} {\bibfnamefont {U.}~\bibnamefont {Wieser}},
  \bibinfo {author} {\bibfnamefont {U.}~\bibnamefont {Kunze}}, \bibinfo
  {author} {\bibfnamefont {G.}~\bibnamefont {Abstreiter}},\ and\ \bibinfo
  {author} {\bibfnamefont {K.}~\bibnamefont {Brunner}},\ }\href
  {https://doi.org/10.1016/j.spmi.2004.02.003} {\bibfield  {journal} {\bibinfo
  {journal} {Superlattices Microstruct.}\ }\textbf {\bibinfo {volume} {33}},\
  \bibinfo {pages} {271} (\bibinfo {year} {2003})}\BibitemShut {NoStop}%
\bibitem [{\citenamefont {Tracy}\ \emph {et~al.}(2009)\citenamefont {Tracy},
  \citenamefont {Hwang}, \citenamefont {Eng}, \citenamefont {{Ten Eyck}},
  \citenamefont {Nordberg}, \citenamefont {Childs}, \citenamefont {Carroll},
  \citenamefont {Lilly},\ and\ \citenamefont {{Das Sarma}}}]{Tracy2009}%
  \BibitemOpen
  \bibfield  {author} {\bibinfo {author} {\bibfnamefont {L.~A.}\ \bibnamefont
  {Tracy}}, \bibinfo {author} {\bibfnamefont {E.~H.}\ \bibnamefont {Hwang}},
  \bibinfo {author} {\bibfnamefont {K.}~\bibnamefont {Eng}}, \bibinfo {author}
  {\bibfnamefont {G.~A.}\ \bibnamefont {{Ten Eyck}}}, \bibinfo {author}
  {\bibfnamefont {E.~P.}\ \bibnamefont {Nordberg}}, \bibinfo {author}
  {\bibfnamefont {K.}~\bibnamefont {Childs}}, \bibinfo {author} {\bibfnamefont
  {M.~S.}\ \bibnamefont {Carroll}}, \bibinfo {author} {\bibfnamefont {M.~P.}\
  \bibnamefont {Lilly}},\ and\ \bibinfo {author} {\bibfnamefont
  {S.}~\bibnamefont {{Das Sarma}}},\ }\href
  {https://doi.org/10.1103/PhysRevB.79.235307} {\bibfield  {journal} {\bibinfo
  {journal} {Phys. Rev. B}\ }\textbf {\bibinfo {volume} {79}},\ \bibinfo
  {pages} {235307} (\bibinfo {year} {2009})}\BibitemShut {NoStop}%
\bibitem [{\citenamefont {{Paquelet Wuetz}}\ \emph {et~al.}(2023)\citenamefont
  {{Paquelet Wuetz}}, \citenamefont {{Degli Esposti}}, \citenamefont {Zwerver},
  \citenamefont {Amitonov}, \citenamefont {Botifoll}, \citenamefont {Arbiol},
  \citenamefont {Vandersypen}, \citenamefont {Russ},\ and\ \citenamefont
  {Scappucci}}]{PaqueletWuetz2023}%
  \BibitemOpen
  \bibfield  {author} {\bibinfo {author} {\bibfnamefont {B.}~\bibnamefont
  {{Paquelet Wuetz}}}, \bibinfo {author} {\bibfnamefont {D.}~\bibnamefont
  {{Degli Esposti}}}, \bibinfo {author} {\bibfnamefont {A.-M.~J.}\ \bibnamefont
  {Zwerver}}, \bibinfo {author} {\bibfnamefont {S.~V.}\ \bibnamefont
  {Amitonov}}, \bibinfo {author} {\bibfnamefont {M.}~\bibnamefont {Botifoll}},
  \bibinfo {author} {\bibfnamefont {J.}~\bibnamefont {Arbiol}}, \bibinfo
  {author} {\bibfnamefont {L.~M.~K.}\ \bibnamefont {Vandersypen}}, \bibinfo
  {author} {\bibfnamefont {M.}~\bibnamefont {Russ}},\ and\ \bibinfo {author}
  {\bibfnamefont {G.}~\bibnamefont {Scappucci}},\ }\href
  {https://doi.org/10.1038/s41467-023-36951-w} {\bibfield  {journal} {\bibinfo
  {journal} {Nat. Commun.}\ }\textbf {\bibinfo {volume} {14}},\ \bibinfo
  {pages} {1385} (\bibinfo {year} {2023})}\BibitemShut {NoStop}%
\end{thebibliography}

%

\end{document}